\documentclass[12pt]{article} 
\usepackage{amsfonts}
\usepackage{latexsym}
\usepackage{amsmath,amssymb}
\usepackage{verbatim}
\usepackage{mathtools}
\usepackage{color}
\usepackage[color,matrix,arrow]{xy}
\usepackage{changebar}
\usepackage{setspace}
\usepackage{simplewick}
\usepackage{tocvsec2}
\usepackage{young}
\usepackage[vcentermath]{youngtab}
\usepackage{array}
\usepackage{cite}
\newenvironment{myenumerate}
{\begin{enumerate}
  \setlength{\itemsep}{0pt}
  \setlength{\parskip}{0pt}
  \setlength{\parsep}{0pt}}
{\end{enumerate}}

\numberwithin{equation}{section}

\newcommand{\bea}{\begin{eqnarray}}
\newcommand{\eea}{\end{eqnarray}}
\newcommand{\be}{\begin{equation}}
\newcommand{\ee}{\end{equation}}
\newcommand{\ba}{\begin{align}}
\newcommand{\ea}{\end{align}}

\newcommand{\azb}{a_{\bar{z}}}

  \makeatletter
  \let\over=\@@over \let\overwithdelims=\@@overwithdelims
  \let\atop=\@@atop \let\atopwithdelims=\@@atopwithdelims
  \let\above=\@@above \let\abovewithdelims=\@@abovewithdelims

\DeclareMathOperator{\Tr}{Tr}

\begin{document}

\ \\

\begin{center}

{\LARGE {\bf Modular Properties of 3D Higher Spin Theory}}

\vspace{0.8cm}

{\large Wei Li,$^{a}$ Feng-Li Lin,$^b$ and Chih-Wei Wang$^b$}
\vspace{1cm}

{\it $^a$Max-Planck-Institut f\"ur Gravitationsphysik, Albert-Einstein-Institut, \\
Am M\"uhlenberg 1, 14476 Golm, Germany} \\
 \tt{\small wei.li@aei.mpg.de}

\vspace{0.9cm}

{\it $^b$Department of Physics, National Taiwan Normal University, \\
Taipei, 116, Taiwan \\
\tt{\small linfengli@phy.ntnu.edu.tw}, \tt{\small freeform1111@gmail.com}
 }

\vspace{1.0cm}

\end{center}

\begin{abstract}

In the three-dimensional $\mathfrak{sl}(N)$ Chern-Simons higher-spin theory, we prove that the conical surplus and the black hole solution are related by the S-transformation of the modulus of the boundary torus. Then applying the modular group 
on a given conical surplus solution, we generate a `$\textrm{SL}(2,\mathbb{Z})$' family of smooth constant solutions. We then show how these solutions are mapped into one another by coordinate transformations that act non-trivially on the homology of the boundary torus. 

After deriving a thermodynamics that applies to all the solutions in the `$\textrm{SL}(2,\mathbb{Z})$' family, we compute their entropies and free energies, and determine how the latter transform under the modular transformations. 
Summing over all the modular images of the conical surplus, we write down a (tree-level) modular invariant partition function.

\end{abstract}

\pagestyle{empty}

\pagebreak
\setcounter{page}{1}
\pagestyle{plain}

\setcounter{tocdepth}{2}
\begin{singlespace}
\tableofcontents
\end{singlespace}

\section{Introduction and Summary}\label{sec:intro}

In the theory of the three-dimensional pure Einstein gravity with a negative cosmological constant, as there is no propagating degree of freedom in the bulk, all asymptotically AdS solutions  are locally diffeomorphic and differ only in their global structures \cite{Cangemi:1992my,Banados:1992gq,Carlip:1994gc,Steif:1995zm}.  In Euclidean signature, starting with a thermal AdS$_3$, whose conformal boundary is a torus with modulus $\tau$, we can obtain an `$\textrm{SL}(2,\mathbb{Z})$' family of solutions via modular transformations ($\tau \mapsto \frac{a\tau+b}{c\tau+d}$) on the modulus of the boundary torus 
\cite{Maldacena:1998bw}. In particular, the S-transformation $\tau \mapsto -\frac{1}{\tau}$ maps the AdS$_3$ into a BTZ black hole. The full modular invariant partition function consists of a sum of all the modular images of the AdS$_3$ partition function: $Z[\tau]=\sum Z_{\textrm{AdS}_3}[\frac{a\tau+b}{c\tau+d}]$ \cite{Dijkgraaf:2000fq}. One can then study the phase structure using $Z[\tau]$, and for example show how  Hawking-Page transition \cite{Hawking:1982dh} occurs when $\tau$ moves across the boundary between different fundamental domains \cite{Maldacena:1998bw,Dijkgraaf:2000fq,Maloney:2007ud
}.

Vasiliev's higher-spin theory is a generalization of the gravity theory; besides the graviton, it contains massless spin-$s$ fields with $s\geq 3$; and it lives in spaces with non-zero constant curvatures, i.e. AdS or dS spaces \cite{Vasiliev:1989re,Vasiliev:1995dn,Vasiliev:1999ba, Vasiliev:2000rn}.  In three dimensions, it can be consistently truncated to a Chern-Simons subsector after the scalar in the theory is decoupled \cite{Blencowe:1988gj,Bergshoeff:1989ns}.\footnote{In 3D, the scalar does not sit in the higher-spin multiplet therefore can be consistently decoupled.} We only consider AdS space in this paper. In AdS$_3$, the gauge algebra of this Chern-Simons theory is an infinite-dimensional Lie algebra $\textrm{hs}[\lambda]$ \cite{Prokushkin:1998bq,Prokushkin:1998vn
}, which at $\lambda=N$ reduces to the finite-dimensional $\mathfrak{sl}(N)$ \cite{Vasiliev:1989re,Fradkin:1990qk}. The 3D Chern-Simons high-spin theory is a straightforward generalization of the $\mathfrak{sl}(2)$ Chern-Simons theory (the alternative formulation of the 3D pure gravity with a negative cosmological constant \cite{Achucarro:1987vz,Witten:1988hc}) and share its essential features: in particular, it does not have any propagating degree of freedom in the bulk of the three-manifold $\mathcal{M}$; the topology of $\mathcal{M}$ and the boundary data on $\partial M$ determine the dynamics. 

In the Chern-Simons higher-spin theory, two types of smooth solutions have been found and studied:  the conical surplus constructed by 
\cite{Castro:2011iw} and the black hole  by \cite{Gutperle:2011kf}. They can be viewed as the higher-spin-charge-carrying generalizations of the AdS$_3$ and BTZ black hole, respectively. Since the notion of the boundary torus still exists in the higher-spin version of the Chern-Simons theory, we can ask whether these two solutions, the conical surplus and the black hole, are related via an S-transformation of the modulus of the boundary torus. 

In some sense, this has to happen  since these two solutions reduce to AdS$_3$ and BTZ when all the higher-spin charges are set to zero.  The non-trivial part of the story is this: the boundary modulus $\tau$ can be considered as the thermodynamical conjugate of the spin-$2$ charge, and in the higher-spin theory, the spin-$2$ field is coupled with all the higher-spin fields, therefore a transformation of $\tau$ inevitably induces the corresponding transformations on the chemical potentials of all the higher-spin charges. The crux in generalizing the modular properties of the spin-$2$ theory to a higher-spin theory is in determining this induced transformations on the higher-spin chemical potentials.

For this purpose, one first needs a consistent description of the thermodynamics of the given solution. Up till now this is absent for the conical surplus; however for the black hole an extensive literature on its thermodynamics has already emerged: e.g. a `holomorphic' approach represented by  \cite{Gutperle:2011kf,Ammon:2011nk,Castro:2011fm,David:2012iu,Chen:2012ba,Ferlaino:2013vga
} and a `canonical' one represented by \cite{Perez:2012cf,deBoer:2013gz,Perez:2013xi}.

In this paper we choose the `canonical' formalism because, as we will show later, in this formalism important quantities and equations are  manifestly modular invariant or covariant. In the `canonical' formalism we generalize the thermodynamics of the black hole to include all smooth stationary solutions. 

Once the thermodynamics of the conical surplus is established, we determine the required transformation on the higher-spin chemical potential accompanying the S-transformation on the boundary modulus $\tau \mapsto -\frac{1}{\tau}$, and prove that the conical surplus and the black hole are S-dual. We then show that the black hole and the conical surplus are related by a coordinate transformation (that changes the modular parameter of the boundary torus from $\tau$ to $-\frac{1}{\tau}$).

Then we generalize from the S-transformation to the full modular group: starting with a higher-spin-charge-carrying conical surplus solution and applying on it the full modular group, we can generate an `$\textrm{SL}(2,\mathbb{Z})$' family of smooth stationary solutions. They are all connected by coordinate transformations that act non-trivially on the homology of the boundary torus. Their free energies hence their on-shell partition functions are related via modular transformations. The full modular invariant partition function then involves summing over all the modular images of the conical surplus.

The paper is organized as follows. In Section~\ref{sec:review} we briefly review some basics of 
3D $\mathfrak{sl}(N)$ higher-spin theory, summarize known stationary smooth solutions, and define new smooth stationary solutions in the `$\textrm{SL}(2,\mathbb{Z})$' family. In Section~\ref{sec:thermodynamics} we formulate a thermodynamics that is universal to all members of the `$\textrm{SL}(2,\mathbb{Z})$' family (including conical surplus and black hole).  
Then in Section~\ref{sec:sdual} we prove that a conical surplus can be mapped into a black hole via an S-transformation of the modulus of the boundary torus. In section~\ref{sec:modularfamily} we show how to generate an `$\textrm{SL}(2,\mathbb{Z})$' family of smooth solutions. We summarize and discuss open problems in section~\ref{sec:conclude}. In Appendix~\ref{app:sigmamu} we present a detailed proof for a statement that is central to our paper; in Appendix~\ref{app:spin2} we review the spin-$2$ story; finally in Appendix~\ref{app:sl4} we discuss $\mathfrak{sl}(4)$ theory as a concrete example.

\bigskip
\bigskip

\section{Basics of 3D higher-spin theory}\label{sec:review}

In this section we first review some basics of the three-dimensional $\mathfrak{sl}(N)$ higher-spin theory (for more details see the earlier works \cite{Campoleoni:2010zq,Gutperle:2011kf,Castro:2011iw} and the reviews \cite{Gaberdiel:2012uj, Ammon:2012wc}). Then we summarize known smooth solutions in this theory, i.e. the conical surplus and the black hole, and meanwhile prepare the readers for our later discussion on general smooth solutions. In this paper we focus on stationary, axially symmetric, solutions.

\bigskip
\subsection{Action}

In three dimensions, the Einstein-Hilbert action with a negative cosmological constant $\Lambda$ can be rewritten in terms of an $\mathfrak{sl}(2,\mathbb{R})\oplus \mathfrak{sl}(2,\mathbb{R})$ Chern-Simons theory (up to a boundary term) \cite{Achucarro:1987vz,Witten:1988hc}:
\begin{equation}\label{doubleChernSimon}
S=S_{\textrm{CS}}[A]-S_{\textrm{CS}}[\bar{A}] \qquad \textrm{with }\quad S_{\textrm{CS}}[A]=\frac{k}{4\pi}\int_{\mathcal{M}} \Tr[A \wedge dA +\frac{2}{3}A\wedge A \wedge A]
\end{equation}
where $\mathcal{M}$ is a locally AdS$_3$ manifold; $A$ and $\bar{A}$ are $\mathfrak{sl}(2,\mathbb{R})$ gauge fields; the trace `$\Tr$' is on the $2$-dimensional representation of $\mathfrak{sl}(2)$ and the level $k=\frac{l_{\textrm{AdS}_3}}{4G_N}$. 
This trick (of rewriting Einstein-Hilbert action with a negative $\Lambda$ into a gauge theory) only works in three dimensions. 

On the other hand, the three-dimensional Vasiliev higher-spin theory is also much more tractable than its higher-dimensional siblings. Besides the gauge fields, the theory has only one additional scalar field, which can be consistently decoupled since it is not part of the higher-spin multiplet in 3D. The gauge field subsector can then be written as a Chern-Simons theory (\ref{doubleChernSimon})
with $A$ and $\bar{A}$ $\in \textrm{hs}[\lambda]$,  
which is an infinite-dimensional Lie algebra and at $\lambda=N$ reduces to $\mathfrak{sl}(N)$ (after quotiented by an infinite ideal). In this paper we will focus on the 3D $\mathfrak{sl}(N)$ Chern-Simons theory. The trace `$\Tr$' in the Chern-Simons action (\ref{doubleChernSimon}) is now on the $N$-dimensional representation of $\mathfrak{sl}(N)$ and the level becomes $k=\frac{\ell_{\textrm{AdS}}}{4G_N}\frac{1}{2\Tr[(L_0)^2]}$.\footnote{This additional normalization factor $\frac{1}{2\Tr[(L_0)^2]}$ is necessary for the spin-$2$ subsector of $\mathfrak{sl}(N)$ higher-spin theory to match the Einstein gravity.}

\medskip

Now let us parametrize the base manifold $\mathcal{M}$. Since the theory has a negative cosmological constant, we choose the boundary condition to be asymptotically AdS$_3$. A constant-time slice of an asymptotically AdS$_3$ space $\mathcal{M}$ is topologically a disc. Specify a radial coordinate $\rho$ and an angular coordinate $\phi$, the coordinate is then $\{\rho, t,\phi \}$. The asymptotic boundary $\partial \mathcal{M}$ is at $\rho \to \infty$ and the boundary coordinates are $\{t,\phi\}$.

In this paper, we focus on Euclidean signature. The Wick rotation into Euclidean signature is via $t\mapsto i t_E$ and the coordinate becomes $\{\rho, z,\bar{z} \}$ with
\begin{equation}
z\equiv \phi + i t_E \ .
\end{equation}
Accordingly, the gauge symmetry becomes $\mathfrak{sl}(N,\mathbb{C})$. The connection $A\in \mathfrak{sl}(N,\mathbb{C})$; and in the representation we will choose, $\bar{A}$ is $A$'s anti-hermitean conjugate:\footnote{This is also the convention used by \cite{Banados:1998gg,Castro:2011iw,deBoer:2013gz} }
\begin{equation}\label{antihermitean}
\bar{A}=-A^{\dagger}\ .
\end{equation}
Therefore, most of the time we only need to write the $A$'s side of the expression and the one for $\bar{A}$ can then be inferred using (\ref{antihermitean}).

\medskip

The Chern-Simons action (\ref{doubleChernSimon}) has gauge degrees of freedom $A \sim A+ d\Lambda$, which allows us to fix the gauge as:
\begin{eqnarray}\label{gauge}
A(\rho,z, \bar{z})=b^{-1}\ a(z,\bar{z}) \ b+b^{-1} \ d b 
\end{eqnarray} 
where $b=e^{\rho L_0}$ is an $\textrm{SL}(N)$-valued $0$-form, and $a$ is an $\mathfrak{sl}(N)$-valued $1$-form on the boundary $\partial \mathcal{M}$: 
\begin{equation}
a=a_{z}dz+a_{\bar{z}}d\bar{z}\,.
\end{equation}

\bigskip

\subsection{Asymptotic symmetries}

The 
$\mathfrak{sl}(2)$ subalgebra corresponds to the spin-$2$ (i.e. gravity) sector, with generators $\{L_{0,\pm 1}\}$, whose commutators are:
\begin{equation}
[L_m,L_n]=(m-n)L_{m+n}\,, \qquad m,n=-1,0,1\,.
\end{equation}
(We will also sometimes write $L_n=W^{(2)}_n$.) From the $\mathfrak{sl}(N)$
gauge symmetry, we first need to make a choice as to which $\mathfrak{sl}(2)$ subalgebra 
corresponds to the gravity sector, namely we need to choose how the gravity $\mathfrak{sl}(2)$ embeds in the full gauge algebra $\mathfrak{sl}(N)$. The choice of this embedding then determines the spectrum of the theory. The principal embedding 
is particularly simple because the field of each spin appears once and only once. In this paper we only discuss the principal embedding and a generalization to other embeddings is straightforward.

Next, one can choose the boundary condition for $A$ and determine the asymptotic symmetry group. This was done in \cite{Henneaux:2010xg,Campoleoni:2010zq,Gaberdiel:2011wb,Campoleoni:2011hg}. In the absence of sources, the asymptotic AdS condition implies 
\begin{equation}\label{naivebc}
A_{\bar{z}}=0\,.
\end{equation} 
However this boundary condition (\ref{naivebc}) is too weak and gives rise to a phase space that is too large (with an affine $\mathfrak{sl}(N)$ algebra as its asymptotic symmetry). An additional boundary condition was proposed by \cite{Campoleoni:2010zq} to supplement (\ref{naivebc}) (see also \cite{Coussaert:1995zp} for the spin-2 case):
\begin{equation}\label{bcDS}
(A-A_{AdS})|_{\rho \to \infty}=\mathcal{O}(1) \,,
\end{equation}
which reduces the phase space by imposing a first-class constraint on the 
$\mathfrak{sl}(N)$ 
affine algebra and results in a $\mathcal{W}_{N}$ algebra as the asymptotic symmetry. This is the bulk realization of the Drinfeld-Sokolov reduction (the reduction of an affine algebra to a $\mathcal{W}$-algebra) \cite{Drinfeld:1984qv}.  

In the process of the Drinfeld-Sokolov reduction, different gauge choices give different bases for the $\mathcal{W}$-algebra. A particular convenient choice is the highest-weight gauge, which gives rise to a $\mathcal{W}$-algebra in which all $W^{(s)}$ are primaries with respect to the lowest spins \cite{Campoleoni:2010zq,Campoleoni:2011hg}. Since there is no spin-$1$ field in the $\mathfrak{sl}(N)$ Chern-Simons theory, all $W^{(s\geq 3)}$ fields are Virasoro primaries:
\begin{eqnarray}\label{Vprimary}
[L_m,W^{(s)}_n] &=& [(s-1) m-n] W^{(s)}_{m+n}\,, \qquad s=3,\ldots,N\,, \quad m,n \in \mathbb{Z}\,.
\end{eqnarray}
This is the gauge we will use throughout this paper.\footnote{Other gauges are possible and might be more suitable for other questions, for details see \cite{Campoleoni:2011hg}.} 

Recall that in the spin-$2$ case the bulk isometry 
$\mathfrak{sl}(2)$ 
is given by the `wedge' subalgebra (generated by $L_{-1,0,1}$) of Virasoro algebra. Here the $(N^2-1)$-dimensional bulk isometry $\mathfrak{sl}(N)$ is generated by $W^{(s)}_m$ with $s=2,\ldots,N$ and  $m=-s+1,\ldots,s-1$. 
An explicit representation of (\ref{Vprimary}) for $m\leq |N|$ is \cite{Pope:1989sr}:
\begin{equation}\label{Wexplicit}
W^{(s)}_n=(-1)^{s-n-1}\frac{(s+n-1)!}{(2s-2)!}(\textrm{Adj}_{L_{-1}})^{s-n-1}(L_1)^{s-1}  
\end{equation}
where the adjoint action $\textrm{Adj}_A B =[A,B]$.
Lastly, we will choose a convention in which  
\begin{equation}
\left(L_{m}\right)^{\dagger}=(-1)^m L_{-m}
\end{equation}
which together with (\ref{Wexplicit}) implies $(W^{(s)}_{m})^{\dagger}=(-1)^m W^{(s)}_{-m}$ for $s=2,\ldots,N$. In this convention we have (\ref{antihermitean}).

\bigskip
\subsection{Smooth solutions}

\medskip
\subsubsection{Equations of motion}

The equation of motion of the Chern-Simons action is the flatness condition for $A$:
$F\equiv dA+A\wedge A=0$. 
In the gauge (\ref{gauge}), this translates into
the flatness of $a$:
\begin{equation}\label{aEOM}
f\equiv da+a \wedge a=0\ . 
\end{equation}

\medskip
In this paper we will only consider axially-symmetric, stationary, solutions. For these solutions, $A$ has only $\rho$-dependence. In the gauge (\ref{gauge}) , this means that $a$ is constant, hence throughout this paper we will refer to them as constant solutions (although they can rotate). Their equations of motion (\ref{aEOM}) reduce to
\begin{equation}\label{aEOMcomm}
[a_{z}, a_{\bar{z}}]=0\,. 
\end{equation}
Once $a_{z}$ is fixed, $a_{\bar{z}}$ can be determined via the equation of motion (\ref{aEOMcomm}), whose solution is simply $a_{\bar{z}}$ being an arbitrary traceless function of $a_{z}$.  By Cayley-Hamilton theorem, an arbitrary function of a $N\times N$ matrix $a_{z}$ (that is generic enough) truncates to a polynomial of $a_z$ of degree-$(N-1)$; therefore $a_{\bar{z}}$ has the expansion \cite{Banados:2012ue}:
\begin{equation}\label{abarz}
a_{\bar{z}}=\sum^{N}_{s=2}\sigma_{s}\left[(a_{z})^{s-1}-\frac{\Tr(a_{z})^{s-1}}{N}\mathbf{1}\right]\,.
\end{equation}
Up to now, $\{\sigma_s\}$ are $(N-1)$ arbitrary complex
parameters. Later we will show how they are fixed in terms of the chemical potentials of the higher-spin charges. 

\medskip
\subsubsection{Holonomy condition}

In this subsection, we define the condition that characterizes a generic smooth constant solution. The known solutions, i.e. the conical surplus and the black hole, are the two special cases.

Any given $a_z$ together with the $a_{\bar{z}}$ related by it via (\ref{abarz}) would solve the equation of motion (\ref{aEOMcomm}). However, requiring the solution be smooth imposes a much more stringent constraint. Since in the higher-spin theory, the spin-$2$ field is coupled with all higher-spin fields hence the Ricci scalar is no longer a gauge invariant entity, the smoothness condition need to be prescribed in terms of other, gauge-invariant, observables. In the 3D Chern-Simons theory, the natural candidate is the holonomy around a one-cycle $\mathcal{C}$ in $\mathcal{M}$:
\begin{equation}
\textrm{Hol}_{\mathcal{C}}(A)\equiv \mathcal{P} e^{\oint_{\mathcal{C}} A}\,.
\end{equation}
The smoothness condition is then simply that the holonomy around any contractible cycle (A-cycle) must be trivial, i.e. $\textrm{Hol}_{\textrm{A}}(A)\in \textrm{center}$ of the gauge group \cite{Castro:2011iw,Gutperle:2011kf}.

First let us describe the cycles in this 3D Euclidean spacetime. The asymptotic boundary of the Euclidean AdS$_3$ is a torus. First we fix its homology basis $(\alpha, \beta)$ with $\alpha \cap \beta=1$. Then we give this torus a complex structure. This allows us to define a holomorphic 1-form $\omega$; we can choose its basis such that $\oint_{\alpha}\omega =1$, then $\tau\equiv \oint_{\beta}\omega$ defines the modulus of the torus. Once the modulus of the boundary torus is fixed, different bulk geometries correspond to different ways of filling the solid torus. We first fix the primitive contractible cycle (A-cycle), then the primitive non-contractible cycle (B-cycle) that satisfies A$\cap$B$=1$ is uniquely determined up to shifts in the A-cycle. The (A,B) homology basis is related to the original $(\alpha, \beta)$ basis via a modular transformation:
\begin{equation}
\begin{pmatrix}
\textrm{B} \\
\textrm{A}
\end{pmatrix}=
\begin{pmatrix}
a &b\\
c& d
\end{pmatrix}  
\begin{pmatrix}
\beta\\
\alpha
\end{pmatrix}\qquad \qquad \textrm{with }\quad
\begin{pmatrix}
a & b\\
c& d
\end{pmatrix} 
\in \textrm{PSL}(2,\mathbb{Z})
\end{equation} with
\begin{equation}
\textrm{PSL}(2,\mathbb{Z})\equiv \textrm{SL}(2,\mathbb{Z})/Z_2\ , \qquad \textrm{SL}(2,\mathbb{Z})\equiv\{\ \begin{pmatrix}
a & b\\
c& d
\end{pmatrix}\ \big\vert \ a,b,c,d \in\mathbb{Z}, ad-bc=1\}\ .
\end{equation}
Then the torus with (A,B) as the homology basis but with the same holomorphic 1-form $\omega$ has a modular parameter
\begin{equation}\label{MPdef}
\textrm{modular parameter}\equiv \frac{\int_{\textrm{B}} \omega}{\int_{\textrm{A}} \omega}=\frac{a \tau+b}{ c \tau +d}\,, \end{equation}
which is a modular transformation of the original modulus $\tau$. Throughout this paper, we use $\gamma$ to denote an element of the modular group $\textrm{PSL}(2,\mathbb{Z})$ and define $\hat{\gamma}\tau$ to be its action on $\tau$:\footnote{In $\textrm{PSL}(2,\mathbb{Z})$, $\gamma$ and $-\gamma$ are identified, hence we can choose $c\geq0$ without loss of generality.}
\begin{equation}\label{taumodular}
\gamma \equiv \begin{pmatrix}
a & b\\
c& d
\end{pmatrix} \in \textrm{PSL}(2,\mathbb{Z}): \qquad \qquad \tau\longmapsto\hat{\gamma}\tau\equiv \frac{a \tau+b}{ c \tau +d}\,.
\end{equation}
Also note that throughout this paper, by \emph{modular parameter} we mean 
the ratio of the (complex) length of the non-contractible(B) cycle and that of the contractible(A) cycle, as defined in 
(\ref{MPdef}); and we reserve the term \emph{modulus} for 
$\tau$. 

\medskip

The conical surplus solutions that carry higher-spin charges has a contractible cycle $\phi \sim \phi+ 2\pi$, just like the AdS$_3$ \cite{Castro:2011iw}.
\begin{equation}
\textrm{CS: } \qquad \gamma= \begin{pmatrix}
1 & 0\\
0& 1 
\end{pmatrix}\quad \Longrightarrow\qquad
\begin{aligned} &\textrm{A-cycle:} \quad z\sim z+ 2 \pi \ , \qquad \\
&\textrm{B-cycle:} \quad z\sim z+ 2 \pi \tau \ .
\end{aligned}
\end{equation} 
Accordingly, the holonomy around this $\phi$-cycle needs to lie in the center of the gauge group: 
\begin{equation}\label{hol-c-1}
\textrm{Hol}_{\phi}(A)=b^{-1} e^{2 \pi \omega_{\phi}} b \ \in \ \textrm{center of }G
\end{equation}
where $\omega_{\phi}$ is defined as\footnote{Throughout the paper we will call such $\omega$ `holonomy matrix'.} 
\begin{equation}\label{omegaphidef}
\omega_{\phi}\equiv  a_{z}+a_{\bar{z}}\,. 
\end{equation}
Let's  denote by `$\Lambda\left(\omega\right)$' the vector of eigenvalues of a matrix $\omega$:
\begin{equation}\label{Spec}
U\  \omega \ U^{-1}=\textrm{DiagonalMatrix}[\lambda_1,\ldots,\lambda_N] \quad \Longrightarrow \quad \Lambda\left(\omega\right)\equiv \vec{\lambda}=(\lambda_1,\ldots,\lambda_N)\,.
\end{equation}
The center of $\textrm{SL}(N,\mathbb{C})$ is $e^{-2\pi i \frac{m}{N}}\mathbf{1}$ with $m\in\mathbb{Z}_N$, which  implies that the eigenvalues of $\omega_{\phi}$ satisfies \cite{Castro:2011iw}:
\begin{equation}\label{holphi}
\Lambda\left({\omega}_{\phi}\right)=i\ \vec{n} \,,
\end{equation}
where the vector $\vec{n}= (n_1,\ldots,n_N)$ with $n_i \in  \mathbb{Z}-\frac{m}{N}$, $n_i\neq n_j$ for $i\neq j$, and $\sum_i^N n_i =0$, and most importantly $n_i$ must come in pairs for the solution to be a conical surplus, i.e. if we order $\{n_i\}$ into a monotone sequence then
\begin{equation}\label{strong}
 n_i+n_{N+1-i}=0\ .
\end{equation}
This imposes a very strong constraint on $m$: $m=0$ for $N$ odd, and $m=0$ or $\frac{N}{2}$ for $N$ even. 
But recall that the center of $\textrm{SL}(N,\mathbb{R})$ is precisely $\mathbf{1}$ for $N$ odd and $\pm\mathbf{1}$ for $N$ even;\footnote{The $\pm \mathbf{1}$ for $N$ even arises from the fact that the gauge group is actually $(\textrm{SL}(N,\mathbb{R})/Z_2)\times (\textrm{SL}(N,\mathbb{R})/Z_2)$ \cite{Castro:2011iw}.} therefore the constraint (\ref{strong}) forces the holonomy to lie in the center of the Lorentzian gauge group $\textrm{SL}(N,\mathbb{R})$ rather than that of the Euclidean one $\textrm{SL}(N,\mathbb{C})$.  In summary the vector $\vec{n}$ obeys
\begin{equation}\label{ndef}
\begin{aligned}
\vec{n} =  (n_1,\ldots,n_N)\,, \quad  n_i \in \begin{cases} \mathbb{Z} &  \ N\mbox{ odd} \\ 
 \mathbb{Z} \ \textrm{or} \ \mathbb{Z}+\frac{1}{2}&  \ N\mbox{ even} 
\end{cases} , \quad n_i\neq n_j \ \textrm{for } i\neq j\, , \quad n_i+n_{N+1-i}=0,
\end{aligned}
\end{equation}
and can be considered as a `topological charge' of the solution; and we will term it `holonomy vector'.
The global AdS$_3$ space corresponds to $ 
\vec{n} =\vec{\rho}$ (the Weyl vector of $\mathfrak{sl}(N)$, with $\rho_i=\frac{N+1}{2} - i$), and generic $\vec{n}$'s satisfying (\ref{ndef}) give conical surpluses \cite{Castro:2011iw}. For discussions on the conical surplus in $\textrm{hs}[\lambda]$ Chern-Simons theory see e.g. \cite{Tan:2012xi,Datta:2012km,Hikida:2012eu,Chen:2013oxa,Campoleoni:2013lma, Campoleoni:2013iha}

\medskip
On the other hand, the (Euclidean) black hole has a  contractible cycle $z \sim z+2\pi \tau$. 
\begin{equation}
\textrm{BH: } \qquad \gamma= \begin{pmatrix}
0 & -1\\
1& 0 
\end{pmatrix}\quad \Longrightarrow\qquad 
\begin{aligned}&\textrm{A-cycle:} \quad z\sim z+ 2 \pi \tau\ , \qquad \\
&\textrm{B-cycle:} \quad z\sim z- 2 \pi \ .
\end{aligned}
\end{equation} Accordingly, the trivial holonomy condition is \cite{Gutperle:2011kf}:
\begin{equation} \label{hol-b-1}
\textrm{Hol}_{t}(A)=b^{-1} e^{2\pi\omega_{t}} b\, \in \, \textrm{center of SL}(N,\mathbb{R})  \,, 
\end{equation}
with $\omega_{t}$ defined as
\begin{equation}\label{omegataudef}
\omega_{t} \equiv \tau a_{z}+\bar{\tau}a_{\bar{z}}\,.
\end{equation}
Namely 
\begin{equation}\label{holtau}
\Lambda\left({\omega}_{t}\right)=i\ \vec{n} \,,
\end{equation}
with $\vec{n}$ satisfying the same set of conditions as the conical surplus (\ref{ndef}). 
$\vec{n}=\vec{\rho}$ corresponds to the higher-spin-charge-carrying BTZ black hole first constructed in \cite{Gutperle:2011kf}; and other $\vec{n}$'s give more generic higher-spin black holes.

\medskip

This definition of smooth solution by the holonomy around the contractible cycle can be easily generalized to include solutions whose modular parameter is a generic $\hat{\gamma}\tau$ (other than $\tau$ or $-\frac{1}{\tau}$). For a generic $\gamma = \tiny\begin{pmatrix}
a & b\\c& d
\end{pmatrix}$ $\in \textrm{PSL}(2,\mathbb{Z})$, the A/B cycles are
\begin{equation}\label{ABcycle}
\begin{aligned}
\textrm{Contractible(A)-cycle: }&& \qquad  z\sim z + 2 \pi (c \tau +d) \,, \\
\textrm{Non-contractible(B)-cycle: } &&\qquad z \sim z + 2 \pi  (a \tau+b)\,.
\end{aligned}
\end{equation}
Accordingly, for a smooth solution, the holonomy around the A-cycle should be trivial
\begin{equation}\label{hologamma1}
\textrm{Hol}_{\textrm{A}}(A)=b^{-1} e^{2\pi\omega_{\textrm{A}}} b \, \in  \,\textrm{center of SL}(N,\mathbb{R})\,,
\end{equation}
with the holonomy matrix around the A-cycle given by
\begin{equation}\label{omegaA}
\omega_{\textrm{A}}\equiv \frac{1}{2\pi}\oint_{\textrm{A}} a=(c\tau+d) a_{z}+(c\bar{\tau}+d)a_{\bar{z}}\,,
\end{equation}
namely 
\begin{equation}\label{holA}
\Lambda\left({\omega}_{A}\right)=i\ \vec{n} 
\end{equation}
again with $\vec{n}$ given by (\ref{ndef}). Here we also write down the holonomy matrix around the B-cycle for comparison and for later use:
\begin{equation}\label{omegaB}
\omega_{\textrm{B}}\equiv \frac{1}{2\pi}\oint_{\textrm{B}} a=(a\tau+b) a_{z}+(a\bar{\tau}+b)a_{\bar{z}}\,.
\end{equation}
For given $\vec{n}$ and $\tau$, varying $\gamma \in \textrm{PSL}(2,\mathbb{Z})$ then generates a `$\textrm{SL}(2,\mathbb{Z})$' family of solutions (a term coined in \cite{Maldacena:1998bw}). Since the T-transformation $\tau \mapsto \tau+1$ does not change the A/B cycle (\ref{ABcycle}), we should consider the subgroup of $\Gamma  \equiv \textrm{PSL}(2,\mathbb{Z})$
\begin{equation}
\Gamma_{\infty}\equiv\{\ \begin{pmatrix}
1 & m\\
0& 1
\end{pmatrix}\ \big\vert \ m\in\mathbb{Z}\}/Z_2 \, \subset\, \Gamma  \equiv \textrm{PSL}(2,\mathbb{Z})  \, .
\end{equation}
to be the stabilizer; hence the `$\textrm{SL}(2,\mathbb{Z})$' family is actually the quotient $\Gamma_{\infty}\backslash\Gamma$. Since a $\gamma$ in $\Gamma_{\infty}\backslash\Gamma$ is uniquely given by the lower row $(c,d)$ (which always satisfies $\textrm{gcd}(c,d)=1$), an enumeration of all the members in this family is thus \cite{serre}:
\begin{equation}
\forall (c,d)  \qquad \textrm{with}\quad c,d\in \mathbb{Z}\,,\quad c\geq0\,,\quad \textrm{gcd}(c,d)=1\,. 
\end{equation}

Lastly, the holonomy vector for $\bar{A}$ is always related to that of $A$ via
\begin{equation}
\vec{\bar{n}}=\vec{n}
\end{equation}
in the anti-hermitean basis we choose. We summarize the discussion of this subsection in the following table:
\begin{table}[!h]
\centering
\begin{tabular}{c | c | c |c }
&EAdS$_3$ and CS  & black hole & Smooth solution $\gamma$\cr
\hline
A-cycle&$z \sim z+ 2\pi$&$z\sim z+2 \pi \tau$ & $z\sim z+2 \pi (c\tau+d)$\cr
\hline
B-cycle  &$z\sim z+2 \pi \tau$&$z \sim z- 2\pi$ &$z\sim z+2 \pi (a\tau+b)$ \cr
\hline
modular parameter&$\tau$&$-\frac{1}{\tau}$ & $\frac{a\tau+b}{c\tau+d}$\cr
\hline
A-cycle holonomy &$\omega_{\phi}=a_{z}+a_{\bar{z}}$& $\omega_{t}=\tau a_{z}+\bar{\tau}a_{\bar{z}}$ &$\omega_{\textrm{A}}=(c\tau+d)a_z+(c\bar{\tau}+d)a_{\bar{z}}$
\end{tabular}
\end{table}


\bigskip
\bigskip

\section{Thermodynamics}\label{sec:thermodynamics}

The conical surplus solution constructed in \cite{Castro:2011iw} has higher-spin charges but with no chemical potential turned on, and hence there is no study on its thermodynamics yet. Meanwhile, the thermodynamics of the black hole in the $\mathfrak{sl}(N)$ Chern-Simons theory has been extensively studied, and depending on the choice of the spin-$2$ conserved charge (the zero-mode of the energy-momentum tensor) there are two main approaches. In the `holomorphic' formalism (initiated in \cite{Gutperle:2011kf} and used in \cite{Ammon:2011nk, Castro:2011fm,David:2012iu,Chen:2012ba,Ferlaino:2013vga}), the spin-$2$ conserved charge (for the left-mover $A$) $T$ is holomorphic.\footnote{For the CFT computation in this formalism see \cite{Kraus:2011ds,Gaberdiel:2012yb,Gaberdiel:2013jca}.} In the `canonical' formalism, the spin-$2$ conserved charge $T$ is obtained either via a canonical approach \cite{Perez:2012cf,Perez:2013xi} \`a la Regge-Teitelboim \cite{Regge:1974zd}, or via a direct derivation from the variational principle \cite{deBoer:2013gz} (for a precursor see \cite{Banados:2012ue}) which gives the same result; in this formalism $T$ is not holomorphic and receives contribution from the right-mover $\bar{A}$. 
The different definitions of $(T,\bar{T})$ in turn leads to different results for the integrability condition,  the entropy, and finally the free energy. For more details see the discussion in \cite{deBoer:2013gz}. (For other discussions on the black hole thermodynamics see \cite{Campoleoni:2012hp,Compere:2013gja}.)

Now we would like to generalize the result of the thermodynamics of the black hole to the conical surplus and to all smooth constant solutions in the `$\textrm{SL}(2,\mathbb{Z})$'
family. Which of the two formalism is better suited for this purpose? Usually the modularity (w.r.t. $\textrm{PSL}(2,\mathbb{Z})$) requires the holomorphicity of the theory. Therefore naively one would expect that the `holomorphic' formalism be the choice whereas the modular property be absent or at least obscured in the `canonical' formalism.

However, the fact that in the `canonical' formalism $T$ lacks holomorphicity therefore modularity does not pose any problem in a discussion of the modular properties in this formalism. First of all, although the spin-2 conserved charge $T$, and hence the entropy, is not modular covariant, this is to be expected since they are not holomorphic to start with. Moreover they are only intermediate quantities. As we will show presently, all the other final quantities --- the connection, the holonomy condition, and the free energy --- are modular invariant or covariant. We will derive manifestly modular covariant expressions for them. We will also show later  in Section~\ref{sec:conclude} that the crucial consistency condition --- the integrability condition (relating conserved charges of different spins) --- is modular invariant in the `canonical' formalism. 

On the other hand, as we will discuss in Section~\ref{sec:conclude}, in the `holomorphic' formalism, it is not clear to us how to write down a simple modular transformation such that the  various important relations --- the holonomy condition, the integrability condition, etc --- are modular invariant or covariant. 

This leads us to choose the `canonical' formalism developed in \cite{deBoer:2013gz} for our generalization of thermodynamics of the black hole to all members of the `$\textrm{SL}(2,\mathbb{Z})$' family. In this section, we generalize the procedure in \cite{Banados:2012ue,deBoer:2013gz}
\begin{myenumerate}
\item Vary the bulk action and identify the source and charge terms in the connection.
\item Write down the suitable boundary action to ensure the variational principle.
\item Identify the conjugate pair of energy and temperature, compute the free energy and entropy, and check the first law of thermodynamics.
\end{myenumerate}
to generic smooth constant solutions (including the conical surplus).
In particular, we compute the on-shell action for generic solutions and write down the modular covariant expressions for the entropy and free energy.

\medskip
One clarification: the construction of these general smooth constant solutions will only be shown later, in Section~\ref{sec:modularfamily}. However, since we first need to know the thermodynamics of the conical surplus solution (in order to consistently turn on its chemical potentials) before we can discuss its relation with the black hole solution, and since the thermodynamics of all these smooth solutions can be discussed in an unified way (and with no need to know the full details of the solutions), we will study all of them at once now, and postpone the explicit construction of these solutions to Section~\ref{sec:modularfamily}.

\bigskip

\subsection{Variational principle}

In the presence of higher-spin conserved charges $Q_s$ with $s=3,\ldots,N$, the partition function (evaluated as an Euclidean path-integral) is a function of the boundary modulus $\tau$ and the chemical potential $\mu_s$ conjugate to the higher-spin charge $Q_s$:
\begin{equation}\label{Zbulk}
Z\left[\tau;\ \mu_s\right]\equiv \int DA D\bar{A} \, e^{-I^{(\textrm{E})}}
\end{equation}
The free energy of the system is
\begin{equation}
-\beta F\left[\tau;\ \mu_s\right]=\ln Z\left[\tau;\ \mu_s\right]\,.
\end{equation}
In this paper, we take the saddle point approximation (i.e. only include the classical result): each classical solution contributes $e^{-I^{(\textrm{E})}\vert_{\textrm{on-shell}}}$. For each classical solution, its free energy $F$ (in the saddle point approximation) is given by
\begin{equation}
-\beta F =-I^{(\textrm{E})}\vert_{\textrm{on-shell}}\,.
\end{equation}
In this section, we study the on-shell action of individual solutions; we will discuss the contributions from all saddle points to the partition function later in Section~\ref{sec:modularfamily}.

\subsubsection{Variation of bulk action}

For the discussion in this section, it is enough to know that a smooth constant solution can be defined by the condition that its holonomy around the A-cycle is trivial, i.e. equation (\ref{ABcycle}), (\ref{hologamma1}), and (\ref{holA}); and it is determined by the $\textrm{PSL}(2,\mathbb{Z})$ element $\gamma$.

The thermodynamics for the case of $\gamma= \tiny\begin{pmatrix}
0 & -1\\
1& 0 
\end{pmatrix}$ (the higher-spin black hole with modular parameter $-\frac{1}{\tau}$) was already given in \cite{deBoer:2013gz}. We now generalize its derivation to the generic smooth solution with modular parameter $\gamma=\tiny\begin{pmatrix}
a & b\\
c& d 
\end{pmatrix}$.
The thermodynamical relation comes out of a direct variational calculation of the Chern-Simons action, which tells us how to add the boundary term once the choice of source/field is made. 
In this variational calculation, the modulus of the boundary torus should actively vary since it carries the physical information of the inverse temperature (and the twist along the angular direction) \cite{Kraus:2006wn,deBoer:2013gz}. However, in the coordinate system $(z,\bar{z})$ which we have been using, the modular parameter $\hat{\gamma}{\tau}$ is hidden in the identification of the A/B cycle; to make it appear explicitly we need to first switch to a coordinate system $(w,\bar{w})$ that lives on a rigid torus with fixed modulus $\tau=i$. The coordinate transformation from $(z,\bar{z})$ to $(w,\bar{w})$ is
\begin{equation}\label{ztow}
z=(c\tau+d) (\frac{1-i\, \hat{\gamma}\tau}{2}w+\frac{1+i\,\hat{\gamma}{\tau}}{2}\bar{w})\end{equation}
and similarly for $\bar{z}$. The A/B cycle is mapped to 
\begin{equation}\label{ABcyclew}
\begin{aligned}
\textrm{A-cycle: }& \qquad z\sim z + 2 \pi (c \tau +d)   \qquad \longmapsto \qquad w\sim w+ 2 \pi\\
\textrm{B-cycle: }& \qquad z \sim z + 2 \pi  (a \tau+b) \qquad \longmapsto \qquad w\sim w+ 2 \pi i
\end{aligned}
\end{equation}
Since in the $(w,\bar{w})$ coordinate the torus is rigid, the on-shell variation of the Euclidean Chern-Simons action 
\begin{equation}
I^{(\textrm{E})}_{\textrm{bulk}}[A]=\frac{ik}{4 \pi}\int \Tr[A\wedge dA +\frac{2}{3}A\wedge A\wedge A]
\end{equation}
in this coordinates is simply
\begin{eqnarray}\label{bulkvariationw}
\delta I^{(\textrm{E})}_{\textrm{bulk}}[A]\vert_{\textrm{on-shell}}=-\frac{ik}{4 \pi}\int_{\partial M} \Tr[a\wedge\delta a]=-\frac{ik}{4 \pi}\int_{\partial M} dw\wedge d\bar{w} \Tr[a_{w}\delta a_{\bar{w}}-a_{\bar{w}}\delta a_{w}]
\end{eqnarray}
and similarly for the $\bar{A}$ term. From now on we will omit the superscript (\textrm{E}) since we will only discuss the Euclidean signature.
Now we translate this back to the $(z,\bar{z})$ coordinate. First, the volume element
is
\begin{equation}\label{volumeelement}
idw \wedge d\bar{w}=\frac{i}{\tau_2}dz\wedge d\bar{z}
\end{equation}
which is invariant under the modular transformation $\tau \mapsto \hat{\gamma}{\tau}$.
Second, the 1-form $a$ is invariant under the coordinate transformation (\ref{ztow}), hence 
\begin{equation}
a_w=(c\tau+d)(\frac{1-i \, \hat{\gamma}{\tau}}{2}a_z)+(c\bar{\tau}+d)(\frac{1-i \,\overline{\hat{\gamma}{\tau}}}{2}a_{\bar{z}})
\end{equation}
whose variation contains explicitly the $\delta \tau$ and $\delta\bar{\tau}$ term:
\begin{equation}
\delta a_{w}= (c\tau+d)(\frac{1-i \, \hat{\gamma}{\tau}}{2}\delta a_z)+(c\bar{\tau}+d)(\frac{1-i \,\overline{\hat{\gamma}{\tau}}}{2}\delta a_{\bar{z}})+\frac{c-i\,a }{2}(a_z\delta \tau  +a_{\bar{z}}\delta \bar{\tau})
\end{equation} 
and similarly for $a_{\bar{w}}$. The integrand in (\ref{bulkvariationw}) is thus
\begin{equation}\label{integrandgamma}
\Tr[a_{w}\delta a_{\bar{w}}-a_{\bar{w}}\delta a_{w}]\Bigr|_{\hat{\gamma}\tau}=\tau_2 \Tr[a_{z}\delta a_{\bar{z}}-a_{\bar{z}}\delta a_{z}]+\frac{i}{2}\Tr[(a_z+a_{\bar{z}})(a_z\delta \tau  +a_{\bar{z}}\delta \bar{\tau})]
\end{equation}
with
\begin{equation}
\tau_2\equiv \textrm{Im} \ \tau
\end{equation}
However this is identical to the corresponding expression for the special case of the modular parameter being $-\frac{1}{\tau}$, i.e. the black hole (see Eq. (4.16) in \cite{deBoer:2013gz}). Namely, the variation of the bulk action only depends on the modulus $\tau$ of the boundary torus but not on $\gamma$, i.e. not on the identification of the A/B cycles. This is to be expected: 
for the theory with boundary torus of the modular parameter $\hat{\gamma}{\tau}$, the inverse temperature $\beta$ and the angular velocity $\theta$ is still given by the modular $\tau$:
\begin{equation}
\frac{\theta+i\beta}{2\pi}=\tau
\end{equation}
instead of $\hat{\gamma}{\tau}$. Accordingly, the zero modes of the stress tensor $(T,\bar{T})$ should still be read off from the coefficients of $(\delta \tau, \delta \bar{\tau})$, instead of $(\delta(\hat{\gamma}\tau), \delta (\hat{\gamma}\bar{\tau}))$. The above computation confirms this.

\medskip

\subsubsection{Charges and chemical potentials}

The fact that the variation of the bulk action (\ref{integrandgamma}) actually does not explicitly depend on $\gamma$ means that all the solutions in this `$\textrm{SL}(2,\mathbb{Z})$' family should have the same identification of source/charge term inside the connection $(a,\bar{a})$. Since for later proofs we will be using some of the details on how $(a,\bar{a})$ depends on the charge and the chemical potential, we will explain them again in detail here, instead of merely referring to the earlier paper \cite{deBoer:2013gz}.

The symmetry algebra $\mathfrak{sl}(N)$ is $(N^2-1)$-dimensional, and can be spanned by the generators $W^{(s)}_m$ with $s=2,\ldots,N$ and $m=-s+1,\ldots,s-1$. The construction (\ref{Wexplicit}) produces an orthogonal basis
\begin{equation}
\Tr[W^{(s)}_m W^{(t)}_n]=t^{(s)}_m\delta^{s,t} \ \delta_{m+n,0}
\end{equation}
where $t^{(s)}_m\equiv \Tr[W^{(s)}_{m}W^{(s)}_{-m}]$ is the normalization factor of $W^{(s)}_{m}$. 


The additional boundary conditions (\ref{bcDS}) are first-class constraints, using which we can bring $a_{z}$ into a form where the charge matrix $\mathbf{Q}$ sits in the highest-weight direction $W^{(s)}_{-s+1}$ \cite{Balog:1990mu}:
\begin{equation}\label{azdef}
a_{z}=L_{1}+\mathbf{Q} \qquad \qquad\mathbf{Q}= \sum^N_{s=2} \frac{Q_s}{t^{(s)}} W^{(s)}_{-s+1}
\end{equation}
where $Q_s$ is the zero mode of the spin-s field $W^{s}$ and 
$t^{(s)}\equiv t^{s}_{-s+1}$. 
Note that in this basis 
\begin{equation}
Q_2=\frac{1}{2}\Tr\left[(a_z)^2\right] \qquad\qquad Q_3=\frac{1}{3}\Tr\left[(a_z)^3\right] 
\end{equation}
but for spins $s\geq 4$, $\frac{1}{s}\Tr\left[(a_z)^s\right]$ is no longer simply $Q_s$, e.g.
\begin{equation}
\frac{1}{4}\Tr\left[(a_z)^4\right]=Q_4+r_1 \ Q^2_2 \qquad \qquad\frac{1}{5}\Tr\left[(a_z)^5\right]=Q_5+r_2 \ Q_2 Q_3 \qquad \ldots
\end{equation}
where $r_{i}$'s are some $N$-dependent rational numbers that can be computed using (\ref{Wexplicit}), e.g. for $N=5$, $r_1=\frac{17}{50}$ and $r_2=\frac{31}{35}$, etc.
Nevertheless we choose this basis because the modular transformation of the connection $(a,\bar{a})$ takes very simple form in this basis, as we will show presently.

\bigskip
$a_{\bar{z}}$ should contain the information of the chemical potentials $\mu_s$'s. Following \cite{deBoer:2013gz}, we demand that the lowest weight terms  (i.e.\ $W^{(s)}_{s-1}$ terms) of $a_{\bar{z}}$ are linear in $\mu$,  in analogue to the construction of $a_z$ in (\ref{azdef}):
\begin{equation}\label{lowestweight}
a_{\bar{z}}=\mathbf{M}+(\textrm{terms  $\sim W^{(s)}_{m \leq s-2}$}) \qquad \qquad  \mathbf{M}=\frac{i}{2 \tau_2} \sum^N_{s=3} \mu_s W^{(s)}_{s-1}
\end{equation}
First note the absence of $\mu_2$ in the definition of $\mathbf{M}$: since we are now in Euclidean signature, the role of $\mu_2$ should be replaced by the modulus $\tau$ of the boundary torus. The prefactor $\frac{1}{2\tau_2}$ is chosen to normalize the $\mu_s Q_s$ term in the on-shell action (see (\ref{variationlnZ})); and it is common to all solutions in the `$\textrm{SL}(2,\mathbb{Z})$' family, as explained in the previous subsection. And the presence of `$i$' is from the Wick rotation to the Euclidean signature.  

From the definition (\ref{lowestweight}) and the orthogonality of the $\{W^{(s)}_m\}$ basis, we have the following $N-1$ equations:
\begin{equation}\label{Nminus1}
\frac{1}{t^{(s)}}\Tr[W^{(s)}_{-s+1} a_{\bar{z}}]=\begin{cases}0&s=2\\
\frac{i}{2 \tau_2} \mu_s\qquad &s=3,\ldots,N
\end{cases}
\end{equation}
which can uniquely determine the $N-1$ $\{\sigma_s\}$ in (\ref{abarz}) in terms of $\{\mu_s\}$ for given $\{Q_s\}$. Once $a_{\bar{z}}$ is written in terms of $\mu_s$ and $Q_s$ we can use the holonomy condition to select the smooth solutions: e.g. (\ref{holphi}) for conical surpluses, (\ref{holtau}) for black holes, and (\ref{hologamma1}) for more generic solutions labeled by $\gamma$. In the canonical ensemble, the higher-spin charges are given and their conjugate chemical potentials are solved in terms of them, the boundary modulus $\tau$,  and the holonomy vector $\vec{n}$ 
\begin{equation}\label{constraintCE}
\mu_t =\mu_t(\vec{n};\ \tau; \ Q_{s\geq 2}) \qquad t=3,\ldots,N
\end{equation}
In the grand canonical ensemble, the chemical potential $\mu_{s\geq 3}$ are given and the charges of the solution are solved in terms of $\mu_s$, the boundary modulus $\tau$,  and the holonomy vector $\vec{n}$ 
\begin{equation}\label{constraintGCE}
Q_t=Q_t(\vec{n};\ \tau; \ \mu_{s \geq 3})  \qquad t=2,\ldots,N
\end{equation}

\subsubsection{Boundary action}

Now that we have made a choice of charge/source terms in $(a_z,a_{\bar{z}})$, we can solve for the boundary action which, when added to bulk action, makes sure that  the variation of the full action has the correct form of [charge $\cdot$ $\delta$ source].

In this derivation, a useful identity is 
\begin{eqnarray}\label{cool1}
\Tr\left[L_{1} a_{\bar{z}}\right]=\Tr\left[ [L_0,\mathbf{Q}] a_{\bar{z}}\right]=\frac{i}{2 \tau_2}\sum^{N}_{s=3}(s-1)\mu_sQ_s 
\end{eqnarray}
where the first `$=$' can be proven using the definition (\ref{azdef}), $[a_z,a_{\bar{z}}]=0$, $[L_0,L_1]=-L_1$, and the cyclic nature of the trace; and the second one is proven by expanding the right hand side of (\ref{cool1}) in terms of (\ref{azdef}) and using $[L_0,W^{(s)}_{-s+1}]=(s-1)W^{(s)}_{-s+1}$. 
And (\ref{cool1}) is also equivalent to 
\begin{equation}\label{thermodynamicsgamma}
\Tr[a_{z}a_{\bar{z}}]=\frac{i}{2 \tau_2} \sum^{N}_{s=3} s\mu_{s}Q_s 
\end{equation}
The appropriate boundary term is
\begin{equation}\label{bndyterm}
I_{\textrm{bndy}}=-\frac{k}{2 \pi}\int_{\partial M} d^2z \Tr[(a_z-2 L_{1}) a_{\bar{z}}]
\end{equation}

Combining the variation of this boundary term with that of the bulk one (eq. (\ref{bulkvariationw}) with (\ref{integrandgamma})), one can first read off the spin-$2$ conserved charges $T$ conjugate to $\tau$
\begin{equation}\label{TTbardef}
\begin{aligned}
T= \frac{1}{2}\Tr\left[(a_{z})^2\right]+\Tr\left[a_z a_{\bar{z}}\right]-\frac{1}{2}\Tr\left[(\bar{a}_{z})^2\right]
\end{aligned}
\end{equation} 
and then write down the variation of the full action (bulk plus boundary): 
\begin{equation}\label{variationlnZ}
\begin{aligned}
&\delta I^{(\textrm{E})} \vert_{\textrm{on-shell}}=\delta I^{(\textrm{E})}_{\textrm{bulk}}\vert_{\textrm{on-shell}}+\delta I^{(\textrm{E})}_{\textrm{bndy}} \vert_{\textrm{on-shell}}\\
=&-(2 \pi i k)\int \frac{d^2z}{4 \pi^2 \tau_2}\left(T\delta\tau-\bar{T}\delta{\tau}+\Tr\left[(a_{z}-L_{1})\delta(-2 i \tau_2 a_{\bar{z}})\right]-\Tr\left[(-\bar{a}_{\bar{z}}+L_{-1})\delta(-2 i \tau_2 \bar{a}_{z})\right]\right)\\
=& -(2 \pi i k) \left(T \ \delta \tau- \bar{T}\ \delta \bar{\tau}+\sum^{N}_{s=3} ( Q_s\ \delta  \mu_s-\bar{Q}_s\ \delta\bar{\mu}_s)\right)
\end{aligned}
\end{equation}
where we have restored the $\bar{A}$ terms. We emphasize that equations (\ref{cool1}) to (\ref{variationlnZ}) have already been given in \cite{deBoer:2013gz} for the black hole case; here we have proved that they are valid for  all solutions in the `$\textrm{SL}(2,\mathbb{Z})$' family:

\bigskip
\subsection{On-shell action, free energy, and entropy}\label{sec:onshell}

For the black hole solution, the free energy and entropy were computed in \cite{Banados:2012ue,deBoer:2013gz}. In this subsection we follow their derivation and determine the entropy and free energy for all the solutions in the `$\textrm{SL}(2,\mathbb{Z})$' family, and more importantly, write them into modular covariant expressions.

As explained earlier, the variation of the action (bulk plus boundary) should take the same form for all solutions in the `$\textrm{SL}(2,\mathbb{Z})$' family, given by (\ref{variationlnZ}).
However the on-shell value of the action depends on how we fill the solid torus therefore is different for different solutions in the  `$\textrm{SL}(2,\mathbb{Z})$' family. \cite{Banados:2012ue} computed the on-shell action of a non-rotating black hole in this $\mathfrak{sl}(N)$ higher-spin theory. Now we generalize to a generic smooth constant solution labeled by $\gamma$. The key point is to choose an appropriate foliation of the three-manifold $\mathcal{M}$ that is regular all the way to the center of $\mathcal{M}$ (i.e. the horizon), thereby avoiding a boundary term at the horizon. In the non-rotating black hole case, the contractible cycle is $t_E\sim t_E+2\pi$, therefore \cite{Banados:2012ue} chose a slicing of the three-manifold $\mathcal{M}$ by disks with constant angular parameter $\phi$,
and obtained an on-shell bulk action $I^{(\textrm{E})}_{\textrm{bulk}}\vert_{\textrm{on-shell}}=-\frac{ik}{4 \pi}\int_{\partial \mathcal{M}}dt d\phi\Tr\left[a_t a_{\phi}\right]$. 

Now for the smooth constant solution labeled by $\gamma$, the contractible cycle is $z\sim z +2 \pi (c\tau+d)$. Go to the coordinate $(u^{\textrm{A}},u^{\textrm{B}})$ defined as
\begin{equation}
u^{\textrm{A}} \equiv \frac{i}{2 \tau_2}\left[(a\bar{\tau}+b)z -(a\tau+b)\bar{z}\right]\,, \qquad \qquad u^{\textrm{B}}\equiv \frac{i}{2 \tau_2}\left[-(c\bar{\tau}+d)z +(c\tau+d)\bar{z}\right]\,.
\end{equation}
We see as $z \mapsto z + 2 \pi (c\tau+d)$,
\begin{equation}
 u^A \longmapsto u^A+2\, \pi\qquad \textrm{and}\qquad u^B \longmapsto u^B\ ;
\end{equation}
i.e. $u^{\textrm{B}}$ comes back to itself around the contractible cycle $z \sim z + 2 \pi (c\tau+d)$. Therefore we should foliate $\mathcal{M}$ with disks of constant $u^{\textrm{B}}$ and such a foliation remains regular inside the bulk; thus we only need to compute the boundary term at the $\rho \rightarrow \infty$. 

Expand the connection $(a,\bar{a}$) in the coordinate $(u^{\textrm{A}},u^{\textrm{B}})$:
\begin{equation}
a=a_{z}dz +a_{\bar{z}}d\bar{z}=\omega_{\textrm{A}} du^{\textrm{A}} + \omega_{\textrm{B}}du^{\textrm{B}} 
\end{equation}
where $\omega_{\textrm{A}}/\omega_{\textrm{B}}$ is precisely the holonomy matrix around the A/B cycle:
\begin{equation}\label{omegaAB}
\omega_{\textrm{A}}=(c\tau+d)a_{z}+(c\bar{\tau}+d)a_{\bar{z}}\ , \qquad \qquad \omega_{\textrm{B}}=(a\tau+b)a_{z}+(a\bar{\tau}+b)a_{\bar{z}}\ .
\end{equation}
Written in term of the coordinate $(u^{\textrm{A}},u^{\textrm{B}})$, the connection $A$ is then
\begin{equation}
\begin{aligned}
A=A_{\rho}d\rho +[ b^{-1} \omega_{\textrm{A}} b]du^{\textrm{A}}+ [b^{-1} \omega_{\textrm{B}} b]du^{\textrm{B}}=A_{\alpha}dx^{\alpha}+A_{\textrm{B}}du^{\textrm{B}}\,,
\end{aligned}
\end{equation}
where $A_{\alpha}dx^{\alpha}=A_{\rho}d\rho+A_{\textrm{A}}du^{\textrm{A}}$ is the projection of $A$ onto the disk (with coordinate $x^{\alpha}=\rho, u^A$); and $b=b(\rho)$ as defined earlier in (\ref{gauge}) --- not to be confused with the entry $b$ in the matrix $\gamma$. The bulk action is thus sliced into 
\begin{equation}\label{bulkaction}
I^{(\textrm{E})}_{\textrm{bulk}}=\frac{ik}{4 \pi}\int du^{\textrm{B}}  \int d\rho du^{\textrm{A}} \Tr\epsilon^{\alpha\beta}\left[A_{\textrm{B}}F_{\alpha\beta}-A_{\alpha}\partial_{\textrm{B}}A_{\beta}\right] -\frac{ik}{4 \pi}\int_{\rho\rightarrow \infty}du^{\textrm{A}} du^{\textrm{B}} \Tr\left[\omega_{\textrm{A}}\omega_{\textrm{B}}\right]\,, 
\end{equation}
with a bulk integral plus a boundary one. 

Now let's compute the on-shell value of (\ref{bulkaction}) for a smooth constant solution labeled by $\gamma$. First, the bulk integral vanishes when taken on-shell, following from the bulk equation of motion $F=0$ and the fact that the solution is constant (i.e. has no $(z,\bar{z})$-dependence). 
Second, in the boundary integral, since the integrand has no dependence on $(z,\bar{z})$,
the integration merely produces an overall volume of the boundary torus $\textrm{vol}(\partial \mathcal{M})=4 \pi^2$.\footnote{The volume element of the boundary torus  is $du^{\textrm{A}} \wedge du^{\textrm{B}}=\frac{i}{\tau_2}dz \wedge d\bar{z}$.}
Therefore restoring the right-movers we conclude that the on-shell (Euclidean) bulk action for the solution $\gamma$ is 
\begin{equation}\label{bulkos}
I^{(\textrm{E})}_{\textrm{bulk}}\vert_{\textrm{on-shell}}=-(2 \pi i k)\frac{1}{2}\Tr\left[\omega_{\textrm{A}} \omega_{\textrm{B}}-\bar{\omega}_{\textrm{A}}\bar{\omega}_{\textrm{B}}\right]\,.
\end{equation}

Then as explained earlier, different solutions in the `$\textrm{SL}(2,\mathbb{Z})$' family share the same boundary term (\ref{bndyterm}). The on-shell value of this boundary action is
\begin{equation}\label{bndyos}
I^{(\textrm{E})}_{\textrm{bndy}}\vert_{\textrm{on-shell}}=(2 \pi i k)\frac{1}{2}\sum^{N}_{s=3} (s-2)(\mu_s Q_s-\bar{\mu}_s\bar{Q}_s)\,.
\end{equation}
Combining (\ref{bulkos}) and (\ref{bndyos}) gives the total on-shell action and hence the free energy:
\begin{equation}\label{freeenergygamma}
\begin{aligned}
-\beta F
=&-(I^{(\textrm{E})}_{\textrm{bulk}}\vert_{\textrm{on-shell}}+I^{(\textrm{E})}_{\textrm{bndy}}\vert_{\textrm{on-shell}})\\
=&(2 \pi i k)\left(\frac{1}{2}\Tr\left[\omega_{\textrm{A}} \omega_{\textrm{B}}-\bar{\omega}_{\textrm{A}}\bar{\omega}_{\textrm{B}}\right] -\frac{1}{2}\sum^{N}_{s=3} (s-2)(\mu_s Q_s-\bar{\mu}_s\bar{Q}_s)\right)\,.
\end{aligned}
\end{equation}

Now let's derive the entropy for generic members of the `$\textrm{SL}(2,\mathbb{Z})$' family. The free-energy should take the form
\begin{eqnarray}\label{freeenergygamma1}
-\beta F&=&S+2 \pi i k\left(T\tau -\bar{T}\bar{\tau} +\sum^{N}_{s=3} (\mu_s Q_s-\bar{\mu}_s\bar{Q}_s)\right) \ .
\end{eqnarray}
We first write down a useful identity
\begin{equation}\label{omegaphiz}
\frac{1}{2}\Tr\left[\omega_{\phi}\omega_{t}-\bar{\omega}_{\phi}\bar{\omega}_{t}\right]  =T\tau-\bar{T}\bar{\tau}+\frac{1}{2}\sum^N_{s=3}s(\mu_sQ_s-\bar{\mu}_s \bar{Q})\,,
\end{equation}
with $\omega_{\phi}$ and $\omega_{t}$ defined earlier in (\ref{omegaphidef}) and (\ref{omegataudef}), respectively.
(\ref{omegaphiz}) can be proven using $(T,\bar{T})$'s definition (\ref{TTbardef}) and the identity (\ref{thermodynamicsgamma}). Using (\ref{omegaphiz}) we obtain the entropy of a smooth constant solution labeled by $\gamma$:
\begin{equation}\label{entropy}
S=(2 \pi i k)\frac{1}{2}\left(\Tr\left[\omega_{\textrm{A}} \omega_{\textrm{B}}-\bar{\omega}_{\textrm{A}}\bar{\omega}_{\textrm{B}}\right]-\Tr\left[\omega_{\phi}\omega_{t}-\bar{\omega}_{\phi}\bar{\omega}_{t}\right]
\right)\ .
\end{equation}

First, let us prove that the entropy defined by (\ref{entropy}) is indeed a Legendre transform of the free energy (\ref{freeenergygamma}). The method is the same as used in \cite{deBoer:2013gz} to prove the corresponding statement for the black hole case. Since the holonomy around the contractible cycle is trivial, the eigenvalue of the holonomy matrix is given by integers hence is rigid, which means
\begin{equation}
\delta \omega_{\textrm{A}}=[\omega_{\textrm{A}},\epsilon] \,,
\end{equation}
where $\epsilon$ is an infinitesimal matrix.
This implies $\Tr\left[a_z \ \delta \omega_{\textrm{A}}\right]=\Tr\left[a_{\bar{z}} \ \delta \omega_{\textrm{A}}\right]=0$, following from $[a_z,a_{\bar{z}}]=0$ for constant solutions. Then using the above together with $\delta\omega_{\textrm{B}}=\frac{a}{c}\delta \omega_{\textrm{A}} -\frac{1}{c}\delta \omega_{\phi}$ we get the variation of the entropy (\ref{entropy}) as: 
\begin{equation}
\delta S=-(2 \pi i k)\Tr\left[\omega_{t}\ \delta \omega_{\phi}-\bar{\omega}_{t}\ \delta \bar{\omega}_{\phi}\right]\,.
\end{equation}
Translated back to the thermodynamical variables $\{T,Q_s;\tau,\mu_s\}$, it is
\begin{equation}
\delta S = -(2 \pi i k) \left(\tau \ \delta T- \bar{\tau}\ \delta \bar{T}+\sum^{N}_{s=3} ( \mu_s\  \delta  Q_s-\bar{\mu}_s\ \delta\bar{Q}_s)\right)\,,
\end{equation}
which confirms that the entropy defined by (\ref{entropy}) is indeed a Legendre transform of the free energy (\ref{freeenergygamma}). 


Finally let's check the modular covariant expression for the entropy in the two special cases, the conical surplus and the black hole. In these two cases, the holonomy matrices around A/B cycle (\ref{omegaAB}) are
\begin{equation}
\begin{aligned}
\textrm{CS: }\qquad & \omega_{\textrm{A}}=a_{z}+a_{\bar{z}}\,, \qquad \qquad \qquad\omega_{\textrm{B}}=\tau a_{z}+\bar{\tau}a_{\bar{z}}\,;\\
\textrm{BH: }\qquad & \omega_{\textrm{A}}=\tau a_{z}+\bar{\tau}a_{\bar{z}}\,, \quad\qquad \qquad \omega_{\textrm{B}}=-a_{z}-a_{\bar{z}}\,.
\end{aligned}
\end{equation}
Since
\begin{equation}\label{omegaABCSBH}
\omega_{\textrm{A}} \omega_{\textrm{B}}=\begin{cases}\omega_{\phi}\omega_{t}  &  \qquad\qquad\textrm{CS} \\ -\omega_{\phi}\omega_{t} & \qquad\qquad\textrm{BH} \end{cases} 
\end{equation}
and similarly for $\bar{\omega}_{\textrm{A}} \bar{\omega}_{\textrm{B}}$,
\begin{equation}\label{SCSBF}
S=\begin{cases} 0 &  \qquad\textrm{CS} \\ -(2 \pi i k)\Tr\left[\omega_{\phi}\omega_{t}-\bar{\omega}_{\phi}\bar{\omega}_{t}\right] =-(2 \pi i k)\left[2(T\tau-\bar{T}\bar{\tau})+\sum^N_{s=3}\ s(\mu_sQ_s-\bar{\mu}_s \bar{Q}_s)\right]
& \qquad\textrm{BH}  \end{cases} 
\end{equation}
Therefore (\ref{entropy}) indeed reproduces the known result for black hole in \cite{deBoer:2013gz} and gives a reasonable answer for the conical surplus. The free energies for these two special cases are thus
\begin{equation}\label{FCSBH}
-\beta F
=(2 \pi i k)\cdot\begin{cases} (T\tau-\bar{T}\bar{\tau})+\sum^N_{s=3}\ (\mu_sQ_s-\bar{\mu}_s \bar{Q}_s)
 &   \qquad\textrm{CS} 
  \\ -(T\tau-\bar{T}\bar{\tau})-\sum^N_{s=3}\ (s-1)(\mu_sQ_s-\bar{\mu}_s \bar{Q}_s)
& \qquad\textrm{BH} \end{cases} 
\end{equation}

\bigskip
\bigskip

\section{Conical surplus and black hole are S-dual}\label{sec:sdual}

In this section, we prove that any given conical surplus solution can be mapped into a black hole under an S-transformation of the modulus $\tau$ of the boundary torus. First, let's state precisely the full action of this S-transformation. 
The S-transformation is the special case of the modular transformation on the boundary torus $\tau \mapsto \hat{\gamma}\tau$ defined in (\ref{taumodular}):
\begin{equation}\label{Strans}
\textrm{S: }\qquad\gamma=
\begin{pmatrix}0& -1\\1 &0\end{pmatrix}: \qquad \qquad \tau \longmapsto \hat{\gamma}\tau= -\frac{1}{\tau}\,.
\end{equation}
Under this S-transformation, the A/B cycles of the conical surplus solution map to those of the black hole:
\begin{equation}\label{ABcycleCStoBH}
\begin{aligned}
\textrm{A-cycle: }&\qquad z \sim z+ 2 \pi   \ \ \qquad \longmapsto \qquad z\sim z+ 2 \pi\tau\\
\textrm{B-cycle: }&\qquad z \sim z + 2 \pi \tau \qquad \longmapsto \qquad z\sim z- 2 \pi 
\end{aligned}
\end{equation}
This would induce the corresponding transformations on the higher-spin charges $Q_s$ and/or their chemical potentials $\mu_s$.

Recall that in the grand canonical ensemble, once the chemical potentials $\{\mu_{s\geq 3}\}$ of a conical surplus are given, the on-shell values of the charges $\{Q_{s\geq 2}\}$ are solved in terms of $\mu_s$, the boundary modulus $\tau$,  and the holonomy vector $\vec{n}$ around its A-cycle (i.e. the $\phi$-cycle):
\begin{equation}\label{chargeCS}
Q^{\textrm{CS}}_t= q_t\left[\vec{n};\ \tau; \ \mu_{s}\right]  \qquad t=2,\ldots,N
\end{equation}
via the trivial holonomy condition around its A-cycle. In this section, we will first show that the S-transformation on the boundary modular parameter (\ref{Strans}) maps the above conical surplus to a black hole with the same chemical potential $\mu_s$, the same holonomy vector $\vec{n}$ but along the new A-cycle (i.e. the $z\sim z+ 2 \pi\tau$ cycle), and with modular parameter $-\frac{1}{\tau}$. The on-shell values of the charges of this black hole are given by
\begin{equation}\label{chargeBH}
Q^{\textrm{BH}}_t=\frac{1}{\tau^t} q_t\left[\vec{n};\ -\frac{1}{\tau}; \ \frac{\mu_{s}}{\tau^s}\right]  \qquad t=2,\ldots,N
\end{equation}
We emphasize that the function $ q_t$ in (\ref{chargeCS}) and (\ref{chargeBH}) is the same one. 

Once (\ref{chargeBH}) is proven, we will then show how the black hole and the conical surplus are related by a coordinate transformation (that changes the modular parameter of the boundary torus from $\tau$ to $-\frac{1}{\tau}$). We then prove that the free energy of the conical surplus and that of the black hole are mapped to each other via the S-transformation (\ref{Strans}). Finally we illustrate with the example of the $\mathfrak{sl}(3)$ theory.

\bigskip
\subsection{S-transformation of holonomy condition and on-shell charges}\label{sec:Sproof}

Since the defining difference between a conical surplus and a black hole is in the holonomy conditions around their respective A-cycles, we will derive the map from (\ref{chargeCS}) to (\ref{chargeBH}) using a map from the holonomy condition of the conical surplus (\ref{holphi}) to that of the black hole (\ref{holtau}).

First let us compare the two holonomy conditions in more details. First, from the equation of motion for constant solutions (\ref{aEOMcomm}), $a_{z}$ and $a_{\bar{z}}$ of a constant on-shell configuration can be simultaneously diagonalized. 
Therefore, the vector of eigenvalues of the holonomy matrix ($\omega_{\phi}$ for the conical surplus and $\omega_{t}$ for the black hole) has a decomposition in terms of vectors of eigenvalues of $a_{z}$ and $a_{\bar{z}}$:
\begin{equation}\label{holonomypairCSBH}
\begin{aligned}
\textrm{CS: }\qquad \Lambda\left(\omega_{\phi}\right)&=\Lambda\left(a_z+a_{\bar{z}}\right)=\Lambda\left(a_z\right)+\Lambda\left(a_{\bar{z}}\right)\,,\\
\textrm{BH: }\qquad \Lambda\left(\omega_{t}\right)&=\Lambda\left(\tau a_z+\bar{\tau}a_{\bar{z}}\right)=\tau \ \Lambda\left(a_z\right)+\bar{\tau}\ \Lambda\left(a_{\bar{z}}\right)\,;
\end{aligned}
\end{equation}
since in computing e.g. $\Lambda\left(a_z+a_{\bar{z}}\right)$ by (\ref{Spec}) we can choose $U$ to be  a unitary matrix that diagonalizes both $a_z$ and $a_{\bar{z}}$.
Recall that  $a_{z}$ is the function of charges $\{Q_{s}\}$ only, as defined in (\ref{azdef}); and $a_{\bar{z}}$ is a function of $\{\mu_s, Q_s\}$ and the modulus $\tau$ as defined via (\ref{abarz}) and (\ref{Nminus1}):\footnote{We emphasize that, just like $a_{z}$, $a_{\bar{z}}$ has exactly the \emph{same} functional form for conical surplus and black hole. The crucial reason is that what is responsible for the thermodynamics is the modulus $\tau$ instead of the modular parameter $\frac{a\tau+b}{c\tau+d}$ in the homology basis of (A-cycle, B-cycle).}  
\begin{equation}
 a_{z}=a_{z}\left[Q_{s}\right]\,,\qquad a_{\bar{z}}=a_{\bar{z}}\left[ \tau; \ \mu_s;\ Q_s \right] \,,
\end{equation}
using which we can write the two holonomy conditions as 
\begin{eqnarray}
\textrm{CS:}&& i\ \vec{n}=\Lambda\left(\omega_{\phi}\left[\tau;\ \mu_s;\ Q_s\right]\right)
=\Lambda\left( a_{z}\left[ Q_s \right] \right)+\Lambda\left( a_{\bar{z}}\left[ \tau; \ \mu_s;\ Q_s\right] \right)\,;\label{defineCS}\\
\textrm{BH:}&& i\ \vec{n}=\Lambda\left(\omega_{t}\left[\tau;\ \mu_s;\ Q_s\right]\right)
=\tau\  \Lambda\left( a_{z}\left[ Q_s\right] \right)+\bar{\tau}\  \Lambda \left( a_{\bar{z}}\left[ \tau;\  \mu_s;\ Q_s  \right] \right)\,.\label{defineBH}
\end{eqnarray}

To further compare the two, we now rewrite (\ref{defineBH}) into the form of (\ref{defineCS}). First, let's look at the $a_z$ part. A crucial observation is that, in the highest-weight gauge, for any $\kappa \in \mathbb{C}$:
\begin{equation}\label{azgeneraltrans}
\kappa \cdot  \kappa^{ L_0}\, a_{z}\left[Q_s\right]\, \kappa^{- L_0}=a_{z}\left[\kappa^{ s} Q_s\right]\,.
\end{equation}
which follows from $a_z$'s definition in the highest-weight gauge (\ref{azdef}) and the commutating relation $[L_0, W^{(s)}_{-s+1}]=(s-1)W^{(s)}_{-s+1}$. Relevant to the present case, we set $\kappa =  \tau$ in (\ref{azgeneraltrans}) and obtain
\begin{equation}\label{aztrans}
\tau \cdot  \tau^{ L_0}\, a_{z}\left[Q_s\right]\, \tau^{- L_0}=a_{z}\left[\tau^{ s} Q_s\right]\,.
\end{equation}
Taking the vectors of eigenvalues of both sides then immediately gives 
\begin{equation}\label{azrescale}
 \tau\  \Lambda\left(a_{z}\left[Q_s\right]\right)=\Lambda\left(a_{z}\left[\tau^s Q_s\right]\right)\,.
\end{equation}

Then we look at the $a_{\bar{z}}$ part. The identity (\ref{azgeneraltrans}) does not have an analogue for $a_{\bar{z}}$, but the special case (\ref{aztrans}) with $\kappa=\tau$ does:
\begin{equation}\label{azbtrans}
 \bar{\tau}\cdot  \tau^{ L_0}\,  a_{\bar{z}}\left[\tau ; \ \mu_s;\ Q_s\right]  \tau^{- L_0} 
 =a_{\bar{z}}\left[ -\frac{1}{\tau};\ \frac{\mu_s}{\tau^s};\ \tau^sQ_s \right] \,.
\end{equation}
Note that from l.h.s. to r.h.s. the variables of $a_{\bar{z}}$ transform as:
\begin{equation}\label{smapfull}
\tau \longmapsto -\frac{1}{\tau}\,, 
\qquad \qquad \mu_s \longmapsto \frac{\mu_s}{\tau^s}\,, 
\qquad \qquad Q_s \longmapsto \tau^sQ_s\,.
\end{equation}

The proof of (\ref{azbtrans}) takes two steps.
\begin{enumerate}
\item Going back to (\ref{abarz}) to rewrite $a_{\bar{z}}$ in terms of only $\sigma_s$ and $Q_s$ (i.e. without the modulus), and then using (\ref{aztrans}), we have
\begin{equation}\label{azbsigmaQ}
\bar{\tau}\cdot\tau^{L_0}\, a_{\bar{z}}\left[\sigma_s; \ Q_s\right]\, \tau^{-L_0}=\sum^N_{s=2}\frac{|\tau|^2\sigma_s}{\tau^{s}} \left[(a_{z}\left[\tau^s Q_s\right])^{s-1} -\frac{\Tr (a_{z}\left[\tau^s Q_s\right])^{s-1}}{N}\right]\,.
\end{equation}
\item Recall that the relations between $\{\sigma_{s\geq 2}\}$ and $\{\tau;\mu_{s\geq 3}\}$ are given by the $(N-1)$ equations in (\ref{Nminus1}). This implies the following identity, which we will prove in Appendix~\ref{app:sigmamu}:
\begin{equation}\label{muQtosigma}
\sigma_s =\frac{i}{2\tau_2}\sum^N_{s^{\prime}=s} \mu_{s^{\prime}} \ H_{s^{\prime}-s} (Q_t)\,,
\end{equation}
where $H_{s^{\prime}-s}(Q_t)$ is a homogenous polynomial (with rational coefficients) of degree-$(s^{\prime}-s)$, with variables $Q_t$ having degree-$t$.
Using that $\tau_2 \mapsto \frac{\tau_2}{|\tau|^2}$ under $\tau\mapsto -\frac{1}{\tau}
$,
we see to get the transformation (\ref{smapfull}) of the variables in (\ref{azbtrans}), we need the following transformation of $(\sigma_s, Q_s)$: 
\begin{eqnarray}
\sigma_s \longmapsto |\tau|^2\frac{\sigma_s}{\tau^s}\,, \qquad\qquad \ Q_s \longmapsto \tau^sQ_s\,.\label{modularsigma}
\end{eqnarray}
Namely, the r.h.s. of (\ref{azbsigmaQ}) is preciely $a_{\bar{z}}\left[ -\frac{1}{\tau};\ \frac{\mu_s}{\tau^s};\ \tau^sQ_s \right]$, which proves (\ref{azbtrans}).

\end{enumerate}

Computing the vectors of eigenvalues of both sides then gives the analogue of (\ref{azrescale}):
\begin{equation}\label{azbarrescale}
 \bar{\tau}\  \Lambda \left(a_{\bar{z}}\left[\tau ; \ \mu_s;\ Q_s\right]\right)
 =\Lambda \left(a_{\bar{z}}\left[ -\frac{1}{\tau};\ \frac{\mu_s}{\tau^s};\ \tau^sQ_s \right]\right) \,.
\end{equation}
Having proved (\ref{azrescale}) and (\ref{azbarrescale}), we can now plug them into the holonomy condition for the black hole (\ref{defineBH}) and recast it into the form of conical surplus' holonomy condition (\ref{defineCS}) but with transformed variables:  
\begin{equation}\label{holoCSBH}
i \ \vec{n}=\Lambda\left(\omega_{t}\left[\tau;\ \mu_s;\ Q_s\right]\right)=\Lambda \left(\omega_{\phi}\left[-\frac{1}{\tau}; \ \frac{\mu_s}{\tau^s};\ \tau^sQ_s \right]\right)\,.
\end{equation}
Equivalently, if we take the holonomy condition of the conical surplus (\ref{defineCS}) and apply on it a change of variables defined in (\ref{smapfull}),
we arrive at the holonomy condition of the black hole (\ref{defineBH}). 

In the proof above, we have adopted the passive viewpoint of transformation and shown that the conical surplus and black hole are mapped to each other via a passive change of variables (\ref{smapfull}). Now we need to translate this into the active viewpoint. To compare the two solutions, we place them into a common grand canonical ensemble: with common temperature and chemical potentials, and ask how their conserved charges are mapped into each other.

In the grand canonical ensemble, once the chemical potentials $\{\mu_{s\geq 3}\}$ are given, the charges $\{Q_s\}$ 
are fixed by the holonomy condition around the A-cycle to be a function of $\{\vec{n};\ \tau;\ \mu_s\}$. The relation (\ref{holoCSBH}) between the holonomy condition of the conical surplus and the black hope then implies the following relation between their 
charge functions:\footnote{Here we emphasize that the relation (\ref{smapcharge}) should not be confused with the last part of the (\ref{smapfull}). The substitution $Q_s \longmapsto \tau^s Q_s$ is part of the passive field redefinition accompanying $\tau \mapsto -\frac{1}{\tau}$; whereas the relation (\ref{smapcharge}) is a bona fide change of the on-shell 
value of the charges (in the grand canonical ensemble). The substitution rule (\ref{smapfull}) implies, rather than contradicts, the map (\ref{smapcharge}). We will discuss this point again from the point of view of the boundary CFT, in Section~\ref{sec:CFT}. 
}
\begin{equation}\label{smapcharge}
Q^{\textrm{CS}}_t= q_t\left[\vec{n};\ \tau; \ \mu_{s}\right] \qquad \Longleftrightarrow \qquad Q^{\textrm{BH}}_t=\frac{1}{\tau^t} q_t\left[\vec{n};\ -\frac{1}{\tau}; \ \frac{\mu_{s}}{\tau^s}\right]  \qquad t=2,\ldots,N\,.
\end{equation}
The function $q\left[\vec{n};\ \tau; \ \mu_{s}\right]$ should be considered as defined using the charge function of the conical surplus solution. 

\bigskip
\subsection{Coordinate transformation between conical surplus and black hole}\label{sec:Scoord}
From the mapping (\ref{holoCSBH}) (or equivalently (\ref{smapcharge})) together with (\ref{aztrans}) and (\ref{azbtrans}), we see that the gauge field components $(a_z,a_{\bar{z}})$ of a conical surplus and those of its S-dual  black hole are related via:
\begin{equation}
\begin{aligned}
a^{\textrm{CS}}_z\left[\vec{n};\, -\frac{1}{\tau};\ \frac{\mu_s}{\tau^s} \right]
&=\tau \cdot \tau^{L_0}\, a^{\textrm{BH}}_z\left[\vec{n};\ \tau;\ \mu_s\right]\,\tau^{-L_0}\,,\\\qquad 
a^{\textrm{CS}}_{\bar{z}}\left[\vec{n};\, -\frac{1}{\tau};\ \frac{\mu_s}{\tau^s} \right]
&=\bar{\tau} \cdot\tau^{L_0} \, a^{\textrm{BH}}_{\bar{z}}\left[\vec{n};\ \tau;\ \mu_s\right]\,\tau^{-L_0} \,.
\end{aligned}
\end{equation}
To further compare the two, let us rescale the coordinate of the conical surplus. Changing the coordinate while fixing the gauge components $(a_{z},a_{\bar{z}})$ gives a different gauge one-form. Under the rescaling of the coordinate $(z,\bar{z})$
\begin{equation}
z\longmapsto z'=\frac{z}{\tau} \qquad\qquad \bar{z}\longmapsto \bar{z}'=\frac{\bar{z}}{\bar{\tau}}\,,
\end{equation}
the gauge one-form $a$ transforms as 
\begin{equation}
a^{(z,\bar{z})}\equiv a_{z}dz+a_{\bar{z}}d\bar{z} \qquad \longmapsto\qquad a^{(z',\bar{z}')}\equiv a_{z}dz'+a_{\bar{z}}d\bar{z}'\,.
\end{equation}
The one-form $a$ of the conical surplus in the new coordinate $(z',\bar{z}')$ is then related to that of the black hole in the original coordinate $(z,\bar{z})$ only by a similarity transformation: 
\begin{equation}\label{Ssimilarity}
 a^{\textrm{CS}}\left[\vec{n};\, -\frac{1}{\tau};\ \frac{\mu_s}{\tau^s} \right]^{(z',\bar{z}')}
 = \tau^{L_0} \, a^{\textrm{BH}}\left[\vec{n};\ \tau;\ \mu_s\right]^{(z,\bar{z})}\, \tau^{-L_0}\,.
 \end{equation}
 Similarly for the right mover:
 \begin{equation}\label{Ssimilaritybar}
\bar{a}^{\textrm{CS}}\left[\vec{n};\, -\frac{1}{\tau};\ \frac{\mu_s}{\tau^s} \right]^{(z',\bar{z}')}
 = \bar{\tau}^{-L_0} \, a^{\textrm{BH}}\left[\vec{n};\ \tau;\ \mu_s\right]^{(z,\bar{z})}\, \bar{\tau}^{L_0}\,.
\end{equation}
Going back to the gauge connection $(A,\bar{A})$, related to $(a,\bar{a})$ via (\ref{gauge}), we see that the similarity transformation in (\ref{Ssimilarity}) and (\ref{Ssimilaritybar}) can be reabsorbed by a simultaneous shifting of the radial coordinate $\rho$ accompanying the rescaling of $(z,\bar{z})$:
\begin{equation}\label{Scoortrans}
\rho\longmapsto \rho'=\rho+\ln |\tau|\,, \qquad z\longmapsto z'=\frac{z}{\tau}\,, \qquad\qquad \bar{z}\longmapsto \bar{z}'=\frac{\bar{z}}{\bar{\tau}}\,.
\end{equation}
Namely, in terms of 
\begin{equation}\label{Anewcoord}
A^{(\rho,z,\bar{z})}\equiv e^{-\rho L_0}a^{(z,\bar{z})} e^{\rho L_0}+L_0 d\rho \qquad\textrm{and}\qquad  A^{(\rho',z',\bar{z}')}\equiv e^{-\rho' L_0}a^{(z',\bar{z}')} e^{\rho' L_0}+L_0 d\rho'
\end{equation}
(and similarly for $\bar{A}$) the full gauge one-form $(A,\bar{A})$ of the black hole with parameter $\{\vec{n};\ \tau;\ \mu_s\}$ in the original coordinate $(\rho,z,\bar{z})$ is identical to that of the conical surplus with parameter $\{\vec{n};\, -\frac{1}{\tau};\ \frac{\mu_s}{\tau^s}\} $ in the new coordinate $(\rho',z',\bar{z}')$ up to an overall constant gauge transformation:
\begin{equation}\label{diffeoS}
\begin{aligned}
\hat{h}^{-1}\cdot A^{\textrm{CS}}\left[\vec{n};\, -\frac{1}{\tau};\ \frac{\mu_s}{\tau^s} \right]
^{(\rho',z',\bar{z}')} \cdot \hat{h}&= A^{\textrm{BH}}\left[\vec{n};\ \tau;\ \mu_s\right]
^{(\rho,z,\bar{z})} \,,\\
\hat{h}^{-1}\cdot \bar{A}^{\textrm{CS}}\left[\vec{n};\, -\frac{1}{\tau};\ \frac{\mu_s}{\tau^s} \right]
^{(\rho',z',\bar{z}')} \cdot \hat{h}&= \bar{A}^{\textrm{BH}}\left[\vec{n};\ \tau;\ \mu_s\right]
^{(\rho,z,\bar{z})} \,,
\end{aligned}
\end{equation}
with $\hat{h}=\left(\frac{\bar{\tau}}{\tau}\right)^{\frac{L_0}{2}}=e^{-i\, \textrm{arg}(\tau) \, L_0}$. This means that the conical surplus and the black hole differ only in the global structure. This is the analogue of the spin-$2$ story: in the spin-$2$ case, the BTZ black hole can be mapped by a coordinate transformation into a thermal AdS$_3$ with modular parameter $-\frac{1}{\tau}$ (instead of $\tau$). They are both discrete quotients of the AdS$_{3}$; thermal AdS$_3$ with parameter $\tau$ whereas BTZ with parameter $-\frac{1}{\tau}$ \cite{Maldacena:1998bw,Dijkgraaf:2000fq}.

\bigskip

\subsection{S-transformation of free energies}\label{sec:smapF}

We have just proved that the S-transformation $\tau\mapsto -\frac{1}{\tau}$ maps a conical surplus with parameter $\{\vec{n};\tau;\mu_s\}$ (holonomy vector around A-cycle, boundary modulus, chemical potentials) into a black hole with the same set of parameters $\{\vec{n};\tau;\mu_s\}$, and that the on-shell value of charges $Q_s$ is mapped via (\ref{smapcharge}). 

Now, between this S-dual pair of a conical surplus and a black hole, if we know the free energy of the conical surplus, as a function of parameters $\{\vec{n};\tau;\mu_s\}$:
\begin{equation}\label{FCS}
F^{\textrm{CS}}=\mathcal{F}\left[\vec{n};\ \tau;\ \mu_s\right]\,,
\end{equation}
what can we say about the free energy of the black hole? In other words, how does the free energy of the conical surplus change under the S-transformation?

To answer this question, we first need to compare the free energies of the two solutions. We have written down a few expressions of  the free energy in Section~\ref{sec:onshell}. For instance the free energies in (\ref{FCSBH}) were written in terms of thermodynamics variables $\{T,Q_s;\, \tau, \mu_s\}$. However, the most important condition in defining an S-dual pair is that the two share the same holonomy vector $\vec{n}$, around their respective A-cycles. Therefore we should instead start with the general expression (\ref{freeenergygamma}), and then use (\ref{omegaABCSBH}) to restrict to the conical surplus or the black hole, then finally rewrite the expression in terms of $\{\vec{n};\ \tau; \, \mu_s;\ Q_s\}$. We now do this separately for the conical surplus and the black hole. 

\medskip
\subsubsection{Conical surplus}

Now let us first rewrite the free energy of the conical surplus in terms of $\{\vec{n};\ \tau; \, \mu_s; \, Q_s\}$.
The general expression for the free energy (\ref{freeenergygamma}) plus (\ref{omegaABCSBH}) give the free energy of the conical surplus to be:
\begin{equation}\label{freeenergyCS}
\begin{aligned}
-\beta F^{\textrm{CS}}
&=(2 \pi i k)\left(\frac{1}{2}\Tr\left[\omega_{\phi} \omega_{t}-\bar{\omega}_{\phi}\bar{\omega}_{t}\right] -\frac{1}{2}\sum^{N}_{s=3} (s-2)(\mu_s Q_s-\bar{\mu}_s\bar{Q}_s)\right)\,,
\end{aligned}
\end{equation}
with $\omega_{\phi}$ and $\omega_{t}$ defined in (\ref{omegaphidef}) and (\ref{omegataudef}), respectively.

First, using 
\begin{equation}
Q_2=\frac{1}{2} \Tr\left[(a_{z})^2\right] \qquad \textrm{ and } \qquad \Tr\left[a_z a_{\bar{z}}\right]= \frac{i}{2\tau_2}\sum^{N}_{s=3}s \mu_s Q_s\,,
\end{equation}
we have
\begin{equation}\label{omegaphizexpand}
\begin{aligned}
\frac{1}{2}\Tr\left[\omega_{\phi}\omega_{t}\right]&=\tau  Q_2 + \tau_1\frac{i}{2\tau_2}\sum^{N}_{s=3}s \mu_s Q_s+ \bar{\tau}\frac{\Tr\left[(a_{\bar{z}})^2\right]}{2} \,, 
\end{aligned}
\end{equation}
where $\tau_1\equiv \textrm{Re}\tau$. The last term in (\ref{omegaphizexpand}) is not immediately defined in terms of the parameters $\{\vec{n};\tau;  \mu_s;  Q_s\}$. However, recall that for the conical surplus, the holonomy matrix $\omega_{\phi}$, with $\Lambda \left(\omega_{\phi}\right)=i\,\vec{n}$ labeling different solutions, contains a $\Tr\left[(a_{\bar{z}})^2\right]$ term: 
\begin{equation}\label{npairCS}
\begin{aligned}
-\frac{1}{2}\vec{n}^2=\frac{1}{2}\Tr\left[(\omega_{\phi})^2\right]&=&Q_2 + \frac{i}{2\tau_2}\sum^{N}_{s=3}s \mu_s Q_s+ \frac{\Tr\left[(a_{\bar{z}})^2\right]}{2}\,.
\end{aligned}
\end{equation}
Therefore (\ref{omegaphizexpand}) and (\ref{npairCS}) together give an identity
\begin{equation}\label{omegaIDCS}
\frac{1}{2}\Tr\left[\omega_{\phi}\omega_{t}\right]^{\textrm{CS}}=-\frac{\vec{n}^2}{2}\bar{\tau}+2 i \tau_2 Q_2-\frac{1}{2}\sum^{N}_{s=3}s \mu_s Q_s\,,
\end{equation}
using which (and its right mover counterpart) we can rewrite the free energy (\ref{freeenergyCS}) of the CS solution  in terms of $\{\vec{n};\ \tau;\ \mu_s;\ Q_s\}$:
\begin{equation}\label{FmodularCS}
-\beta F^{\textrm{CS}}=2\pi i k\left[\ 2 i \tau_2(\frac{\vec{n}^2}{2}+Q_2+\bar{Q}_2)-\sum^{N}_{s=3}(s-1) (\mu_s Q_s-\bar{\mu}_s\bar{Q}_s)\right]\,.
\end{equation}
The set of parameters $\{\vec{n};\ \tau;\ \mu_s;\ Q_s\}$ transform covariantly under modular transformation. However, this is not yet the last step. Since the free energy is in the grand canonical ensemble, we should consider its variables to be $\{\vec{n};\ \tau;\ \mu_s\}$, and replace $Q_s$ by its function of them. Plugging the 
charge function for the conical surplus (\ref{chargeCS}) into (\ref{FmodularCS}), we get the final answer for its free energy
\begin{equation}\label{FmodularCSGC}
\begin{aligned}
-\beta F^{\textrm{CS}}=&2\pi i k\Big[\ 2 i \tau_2\left(\frac{\vec{n}^2}{2}+q_2[\vec{n};\ \tau;\ \mu_s]+\bar{q}_2[\vec{n};\ \tau;\ \mu_s]\right)\\
&-\sum^{N}_{s=3}(s-1) \left(\mu_s q_s[\vec{n};\ \tau;\ \mu_s]-\bar{\mu}_s\bar{q}_s[\vec{n};\ \tau;\ \mu_s]\right)\Big]\equiv -\beta \mathcal{F}[\vec{n};\ \tau;\ \mu_s]\,.
\end{aligned}
\end{equation}
In the last line we have used this result of the free energy of the conical surplus to define a `free energy function' $\mathcal{F}[\vec{n};\ \tau;\ \mu_s]$, which will be used later to relate to the free energy of other solutions in the `$\textrm{SL}(2,\mathbb{Z})$' family.

\medskip
\subsubsection{Black hole}

The black hole case is completely parallel. Applying (\ref{omegaABCSBH}) to the general expression (\ref{freeenergygamma}) gives the free energy of the black hole:
\begin{equation}\label{freeenergyBH}
\begin{aligned}
-\beta F^{\textrm{BH}}
&=(2 \pi i k)\left(-\frac{1}{2}\Tr\left[\omega_{\phi} \omega_{t}-\bar{\omega}_{\phi}\bar{\omega}_{t}\right] -\frac{1}{2}\sum^{N}_{s=3} (s-2)(\mu_s Q_s-\bar{\mu}_s\bar{Q}_s)\right)\,.
\end{aligned}
\end{equation}
The first term involving $\frac{1}{2}\Tr\left[\omega_{\phi} \omega_{t}-\bar{\omega}_{\phi}\bar{\omega}_{t}\right]$ can still be rewritten using (\ref{omegaphizexpand}). But since 
the topological charge $\vec{n}$ is given by the holonomy along the $t$-cycle instead of the $\phi$-cycle, (\ref{npairCS}) should now be replaced by
\begin{equation}\label{npairBH}
\begin{aligned}
-\frac{1}{2}\vec{n}^2=\frac{1}{2}\Tr\left[(\omega_{t})^2\right]
&=&\tau^2 Q_2 + \frac{i|\tau|^2}{2\tau_2}\sum^{N}_{s=3}s \mu_s Q_s+ \bar{\tau}^2\frac{\Tr\left[(a_{\bar{z}})^2\right]}{2}\,, 
\end{aligned}
\end{equation}
which gives rise to the identity (the counterpart of (\ref{omegaIDCS}))
\begin{equation}\label{omegaIDBH}
\begin{aligned}
\frac{1}{2}\Tr\left[\omega_{\phi} \omega_{t}\right]^{\textrm{BH}}&=-\left(\frac{\vec{n}^2}{2}\frac{1}{\bar{\tau}}+\frac{2 i \tau_2}{|\tau|^2} \tau^2 Q_2-\frac{1}{2}\sum^{N}_{s=3}s \mu_s Q_s\right)\,.
\end{aligned}
\end{equation}
Using this the free energy of the black hole can be written in terms of modular covariant quantities $\{\vec{n};\ \tau;\ \mu_s;\ Q_s\}$ as
\begin{equation}\label{FmodularBH}
-\beta F^{\textrm{BH}}=2\pi i k\left[\ \frac{2 i \tau_2}{|\tau|^2}(\frac{\vec{n}^2}{2}+\tau^2 Q_2+\bar{\tau}^2\bar{Q}_2)-\sum^{N}_{s=3}(s-1) (\mu_s Q_s-\bar{\mu}_s\bar{Q}_s)\right]\,.
\end{equation}
Similar to the conical surplus, we should now replace $Q_s$ by the charge functions of the black hole, given in (\ref{chargeBH}), and thereby obtain the free energy of the black hole as a function of $\{\vec{n};\ \tau;\ \mu_s\}$:
\begin{equation}\label{FmodularBHGC}
\begin{aligned}
-\beta F^{\textrm{BH}}=&2\pi i k\Big[\ \frac{2 i \tau_2}{|\tau|^2}\left(\frac{\vec{n}^2}{2}+q_2\left[\vec{n};\ -\frac{1}{\tau}; \ \frac{\mu_{s}}{\tau^s}\right] +\bar{q}_2\left[\vec{n};\ -\frac{1}{\bar{\tau}}; \ \frac{\bar{\mu}_{s}}{\bar{\tau}^s}\right] \right)\\
&-\sum^{N}_{s=3}(s-1) \left(\frac{\mu_s}{\tau^s} q_s\left[\vec{n};\ -\frac{1}{\tau}; \ \frac{\mu_{s}}{\tau^s}\right] -\frac{\bar{\mu}_s}{\bar{\tau}^s}\bar{q}_s\left[\vec{n};\ -\frac{1}{\bar{\tau}}; \ \frac{\bar{\mu}_{s}}{\bar{\tau}^s}\right] \right)\Big]\,.
\end{aligned}
\end{equation}

\medskip
\subsubsection{S-transformation of free energies}

Now we have the free energy of the conical surplus (\ref{FmodularCSGC}) and that of the black hole (\ref{FmodularBHGC}), each written explicitly in terms of $\{\vec{n};\ \tau;\ \mu_s\}$. Comparing (\ref{FmodularCSGC}) with (\ref{FmodularBHGC}), we conclude that the S-transformation 
between the free energy of the conical surplus and that of the black hole is the following 
\begin{equation}\label{SmapF}
F^{\textrm{CS}}= \mathcal{F}\left[\vec{n};\ \tau; \ \mu_{s}\right] \qquad \Longleftrightarrow \qquad F^{\textrm{BH}}=\mathcal{F} \left[\vec{n};\ -\frac{1}{\tau}; \ \frac{\mu_{s}}{\tau^s}\right]  \,.
\end{equation}	
Similar to the charge function (\ref{smapcharge}),  we can consider the function $\mathcal{F}\left[\vec{n};\ \tau; \ \mu_{s}\right]$ to be defined by the free energy of the conical surplus. 
The relations (\ref{smapcharge}) and (\ref{SmapF}) show that the full S-transformation in the grand canonical ensemble (GCE) is:
\begin{equation}\label{Smodulartaumu}
\textrm{S-map in GCE: }\qquad\tau \longmapsto -\frac{1}{\tau}\,, \qquad \qquad \mu_s \longmapsto \frac{\mu_s}{\tau^s}\,.
\end{equation}

\bigskip
\subsection{Example-1: $N=3$ case}
Now let us illustrate what we have proven so far with the simplest example: $\mathfrak{sl}(3)$ higher-spin theory.

First, to write down $a_{z}$ from (\ref{azdef}) we need to specify a specific representation of $L_{0,\pm 1}$. We will adopt the one used in \cite{Castro:2011iw}, namely
\begin{equation}
\begin{aligned}
(L_0)_{mn}&=\frac{N+1-2 m}{2}\ \delta_{m,n}\,,\quad \\
(L_1)_{mn}&=-\sqrt{\vert(m-1)(N+1-m)\vert}\ \delta_{m,n+1}\,,\quad \\
(L_{-1})_{mn}&=+\sqrt{\vert(n-1)(N+1-n)\vert} \ \delta_{m+1,n}\,.
\end{aligned}
\end{equation}
Once $L_{0,\pm 1}$ are chosen, we can use (\ref{Wexplicit}) to write down all the $W^{(s\geq 3)}_m$. In this representation, $a_{z}$ is
\begin{equation}\label{azN3}
a_z =  L_1 - \frac{Q_2}{4} L_{-1} + \frac{Q_3}{4} W^{(3)}_{-2}\,.
\end{equation}
Then take (\ref{abarz}) and solve $\{\sigma_2,\sigma_3\} $ using (\ref{Nminus1}) we get
\begin{equation}
\sigma_{2}=0 \,, \qquad \qquad \sigma_3=\frac{i}{2\tau_2}\mu_3 \,.
\end{equation}
So $a_{\bar{z}}$ is simply
\begin{equation}\label{azbarN3}
\azb = \frac{i}{2\tau_2}\mu_3\left((a_z)^2-\frac{\Tr\left[(a_z)^2\right]}{3}\mathbf{1} \right)\,. 
\end{equation}
Note that for $N\geq 4$, $\sigma_s$ is no longer simply $\frac{i}{2\tau_2}\mu_s$ but in general involves a non-trivial (i.e. $\neq 1$) homogenous polynomial of $Q_s$ as given by (\ref{muQtosigma}). We remind that the expressions of $a_{z}$ and $a_{\bar{z}}$ as given in (\ref{azN3}) and (\ref{azbarN3}) are valid for all members of the `$\textrm{SL}(2,\mathbb{Z})$' family. However the charges $Q_s$ are determined in terms of the chemical potentials via the holonomy condition (\ref{holA}), which is different for different solutions (labeled by $\gamma$).

First let's look at the conical surplus. 
The holonomy condition along the $\phi$-cycle ((\ref{holphi}) and (\ref{omegaphidef})) is equivalent to the following two equations
\begin{equation}\label{hol_sl3CS}
\Tr\left[(a_z + a_{\bar{z}})^2\right] = -2 n^2\,, \qquad \Tr\left[(a_z + a_{\bar{z}})^3\right] =0 \ ,\qquad \textrm{with }n\in \mathbb{Z}
\end{equation}
where we have used $\vec{n}=(n,0,-n)$.\footnote{The solution with chemical potential turned off (i.e. $\mu_3=\bar{\mu}_3=0$) is the conical surplus defined in \cite{Castro:2011iw}. 
} 
Now let us use (\ref{hol_sl3CS}) to solve $\{Q_2,Q_3\}$ in terms of $\tau$ and $\mu_3$. (\ref{hol_sl3CS}) is a pair of coupled algebraic equations, one quadratic and one cubic; therefore $\{Q_2,Q_3\}$ has an algebraic expression in terms of $\{\tau, \mu_3\}$. However they are rather long and not very illuminating so we instead provide (the first few terms of) their power expansion in terms of $\mu_3$, which is enough to show the S-map between this conical surplus solution and the black hole: 
\begin{equation}\label{qcs_sl3}
\begin{aligned}
 Q^{\textrm{CS}}_2 &= -n^2\left[1 - \frac{5}{3}\alpha^2 + \frac{10}{3}\alpha^4  - \frac{221}{27}\alpha^6  +\frac{1802}{81 }\alpha^8 
+ \cdots \right]\equiv q_2[\vec{n};\, \tau;\, \mu_s]\,, \\
 Q^{\textrm{CS}}_3 &=-n^3 \left[\frac{2}{3}\alpha - \frac{40}{27}\alpha^3 + \frac{34}{9}\alpha^5 
- \frac{848}{81}\alpha^7 
+ \cdots\right]\equiv q_3[\vec{n};\, \tau;\, \mu_s]\,.
\end{aligned}
\end{equation}
with
\begin{equation}\label{xCS}
\alpha\equiv n\frac{i}{2\tau_2}\mu_3
\end{equation}
In the last step we defined the function $q_s$ using the charge function of the conical surplus. 
We could then plug the exact solution into (\ref{FmodularCS}) to write down the exact free energy. However, since the expression is too long and not very illuminating, we again content ourselves with an expansion:
\begin{equation}\label{fcs_sl3}
\begin{aligned}
-\beta F^{\textrm{CS}} 
=  4 \pi k \cdot n^2 \tau_2\cdot\left[ 1-\frac{1}{3}(\alpha^2+\bar{\alpha}^2)+\frac{10}{27}(\alpha^4+\bar{\alpha}^4)-\frac{17}{27}(\alpha^6+\bar{\alpha}^6)+\frac{106}{81}(\alpha^8+\bar{\alpha}^8)+\cdots\right]\,.
\end{aligned}
\end{equation}

Now we turn to the black hole. The condition that solves $\{Q_2,Q_3\}$ is now the holonomy condition around the cycle $z \sim z+ 2\pi \tau$:
\begin{equation}\label{hol_sl3BH}
\Tr\left[(\tau a_z + \bar{\tau}a_{\bar{z}})^2\right] = -2 n^2\,, \qquad \Tr\left[(\tau a_z +\bar{\tau} a_{\bar{z}})^3\right] =0 \,,\qquad \textrm{with } \qquad n\in \mathbb{Z}\,.
\end{equation}
The case with $n=1$ gives the higher-spin black hole first constructed in 
\cite{Gutperle:2011kf}. 
Now we write down the solution $\{Q_2,Q_3\}$ for generic $n$, in a power expansion of $\mu_3$:
\begin{equation}\label{qbh_sl3}
\begin{aligned}
Q^{\textrm{BH}}_2 &= -\frac{n^2}{\tau^2}\left[1 - \frac{5}{3}\beta^2 + \frac{10}{3}\beta^4  - \frac{221}{27}\beta^6  +\frac{1802}{81 }\beta^8 
+ \cdots \right]\,, \\
Q^{\textrm{BH}}_3 &=-\frac{n^3}{\tau^3} \left[\frac{2}{3}\beta - \frac{40}{27}\beta^3 + \frac{34}{9}\beta^5 
- \frac{848}{81}\beta^7 
+ \cdots\right]\,.
\end{aligned}
\end{equation}
with
\begin{equation}\label{xBH}
\beta \equiv n\frac{i|\tau|^2}{2\tau_2}\frac{\mu_3}{\tau^3}\,.
\end{equation}
Comparing (\ref{qcs_sl3}) and (\ref{qbh_sl3}) together with (\ref{xCS}) and (\ref{xBH}) thus immediately shows $Q^{\textrm{BH}}_s =\frac{1}{\tau^s}q_{s}\left[\vec{n};\ -\frac{1}{\tau};\ \frac{\mu_s}{\tau^s}\right]$ hence 
confirms the relation (\ref{smapcharge}) that we have proven earlier. Finally we compute the free energy of the black hole and write down its power expansion here
\begin{equation}\label{fbh_sl3}
\begin{aligned}
-\beta F^{\textrm{BH}} 
=  4 \pi k \cdot n^2 \frac{\tau_2}{|\tau|^2}\cdot\left[ 1-\frac{1}{3}(\beta^2+\bar{\beta}^2)+\frac{10}{27}(\beta^4+\bar{\beta}^4)-\frac{17}{27}(\beta^6+\bar{\beta}^6)+\frac{106}{81}(\beta^8+\bar{\beta}^8)+\cdots\right]\,.
\end{aligned}
\end{equation}
This is exactly the S-transformation (\ref{Smodulartaumu}) 
of (\ref{fcs_sl3}).

\bigskip
\bigskip

\section{$\textrm{SL}(2,\mathbb{Z})$ family of smooth solutions}\label{sec:modularfamily}

In this section we will show how to generate an ``$\textrm{SL}(2,\mathbb{Z})$ family"  of smooth solutions with higher-spin charges. The construction is a generalization of the spin-2 case, which we briefly review in Appendix~\ref{app:spin2}.

\bigskip
\subsection{Solution}

In the spin-2 case (i.e. pure Einstein gravity with a negative cosmological constant), all BTZ black holes can be obtained from the AdS$_3$ space via modular transformations of the modulus of the boundary torus (which induces a large coordinate transformation) plus a local coordinate transformation. In the $\mathfrak{sl}(N)$ higher-spin gravity, let us start with a conical surplus solution and apply the full modular group $\textrm{PSL}(2,\mathbb{Z})$.

Under a modular transformation on the boundary torus 
\begin{equation}\label{gamma}
\gamma=\begin{pmatrix}a& b\\c &d\end{pmatrix}\in \textrm{PSL}(2,\mathbb{Z})
: \qquad \qquad \tau \longmapsto \hat{\gamma}\tau= \frac{a\tau+b}{c\tau+d}
\end{equation}
the A/B cycles of the conical surplus solution map to:
\begin{equation}\label{ABcyclegamma}
\begin{aligned}
\textrm{A-cycle: }&\qquad z \sim z+ 2 \pi   \ \ \qquad \longmapsto \qquad z\sim z+ 2 \pi(c\tau+d)\\
\textrm{B-cycle: }&\qquad z \sim z + 2 \pi \tau \qquad \longmapsto \qquad z\sim z+ 2 \pi (a\tau+b) 
\end{aligned}
\end{equation}
The holonomy matrix around the contractible cycle should now be
\begin{equation}\label{holonomygammadef}
\omega_{\textrm{A}}=(c\tau+d)\ a_{z}+(c\bar{\tau}+b)\ a_{\bar{z}}
\end{equation}
which for a smooth solution should be: 
\begin{equation}\label{holonomygamma}
\Lambda\left(\omega_{\textrm{A}}\right)=i \ \vec{n} 
\end{equation}
with $\vec{n}$ satisfying (\ref{ndef}). Let us label this new solution by $\gamma$.

Now, if we start with a given conical surplus with parameter $\{\vec{n};\tau;\mu_s\}$ (holonomy vector around $\phi$-cycle, boundary modulus, chemical potentials), and whose charge function is known to be given by (\ref{chargeCS}), what can we say about this new smooth solution labeled by $\gamma$? Again, since in this theory a solution is defined by the trivial holonomy condition around its A-cycle, the answer must lie in a comparison between the holonomy condition of the conical surplus and that of the solution $\gamma$. Therefore we follow the strategy used earlier (in Section~\ref{sec:Sproof}) in proving the S-duality between the conical surplus and the black hole: we first recast the holonomy condition of the solution $\gamma$ into the form of that of the conical surplus to facilitate the comparison, then we infer the transformation rules from this comparison.

The conical surplus with parameter $\{\vec{n};\tau;\mu_s \}$ is defined by the equation (\ref{defineCS}),  but we now write it again for the ease of comparison:
\begin{equation}\label{defineCS2}
\begin{aligned}
\textrm{CS: }\qquad i\  \vec{n}=\Lambda\left(\omega_{\phi}\left[\tau;\, \mu_s;\, Q_s\right]\right)
=\Lambda\left( a_{z}\left[ Q_s \right] \right)+\Lambda\left( a_{\bar{z}}\left[ \tau; \ \mu_s;\ Q_s\right] \right)\,,
\end{aligned}
\end{equation}
whereas the new smooth constant solution with the same set of parameters $\{\vec{n};\tau;\mu_s\}$ and labeled by $\gamma$ is defined by
\begin{equation}\label{definegamma}
\begin{aligned}
\textrm{$\gamma$: }& \quad i \ \vec{n}=\Lambda\left(\omega_{\textrm{A}}\left[\tau;\, \mu_s;\, Q_s  \right]\right)=(c\tau+d)  \Lambda\left( a_{z}\left[ Q_s\right] \right)+(c\bar{\tau}+d)  \Lambda \left( a_{\bar{z}}\left[\tau; \,\mu_s; \,Q_s  \right] \right)\,.
\end{aligned}
\end{equation}
(We remind that in both cases, $Q_s$ can be solved in terms of $\{\vec{n};\tau;\mu_s\}$, using (\ref{defineCS2}) and (\ref{definegamma}), respectively.)
To cast (\ref{definegamma}) into the form of (\ref{defineCS2}), let us rerun the argument in Section~\ref{sec:Sproof}. 

First, the pair (\ref{aztrans}) and (\ref{azbtrans}) now generalize to 
\begin{equation}\label{transgamma}
\begin{aligned}
(c\tau+d)\cdot  (c\tau+d)^{ L_0}\, a_{z}\left[Q_s\right]\,(c\tau+d)^{- L_0}&=a_{z}\left[(c\tau+d)^{ s} Q_s\right]\,,\\
(c\bar{\tau}+d)\cdot  (c\tau+d)^{ L_0}\,  a_{\bar{z}}\left[\tau ; \ \mu_s;\ Q_s\right]  (c\tau+d)^{- L_0} 
 &=a_{\bar{z}}\left[\hat{\gamma}{\tau};\ \frac{\mu_s}{(c\tau+d)^s};\ (c\tau+d)^sQ_s \right] \,.
\end{aligned}
\end{equation}
The first identity in (\ref{transgamma}) comes from setting $\kappa=c\tau+d$ in (\ref{aztrans}).
To prove the second identity, we again repeat the argument of the S-transformation. First, (\ref{azbsigmaQ}) should now be generalized into  
\begin{equation}\label{azbsigmaQgamma}
\begin{aligned}
&(c\bar{\tau}+d)\cdot(c\tau+d)^{L_0}\, a_{\bar{z}}\left[\sigma_s; \ Q_s\right]\, (c\tau+d)^{-L_0}\\
=&\sum^N_{s=2}\frac{|(c\tau+d)|^2\sigma_s}{(c\tau+d)^{s}} \left[(a_{z}\left[(c\tau+d)^s Q_s\right])^{s-1} -\frac{\Tr (a_{z}\left[(c\tau+d)^s Q_s\right])^{s-1}}{N}\right]\,.
\end{aligned}
\end{equation}
Second, (\ref{muQtosigma}) still applies (note the presence of $\tau_2$ instead of the imaginary part of the modular parameter $\frac{a\tau+b}{c\tau+d}$). The only difference is that, now from l.h.s. to the r.h.s. the variables undergo the following transformation:
\begin{equation}\label{modularfull}
\tau \longmapsto \hat{\gamma}\tau= \frac{a\tau+b}{c\tau+d} \,, \qquad \qquad \mu_s \longmapsto \frac{\mu_s}{(c\tau+d)^s}\,, \qquad \qquad Q_s \longmapsto (c\tau+d)^sQ_s \,,
\end{equation}
of which (\ref{smapfull}) is merely a special case with $\gamma=\tiny\begin{pmatrix}
0 & -1\\
1& 0 
\end{pmatrix}$.
Accordingly, the transformation in terms of $(\sigma_s, Q_s)$ analogous to (\ref{modularsigma}) should be
\begin{equation}
\sigma_s \longmapsto |c\tau+d|^2\frac{\sigma_s}{(c\tau+d)^s}\,, \qquad\qquad \ Q_s \longmapsto (c\tau+d)^sQ_s\,.
\end{equation}
(We have used $\tau_2 \mapsto\frac{\tau_2}{|c\tau+d|^2}$ under $\tau\mapsto \frac{a\tau+b}{c\tau+d}$.)
Namely, the r.h.s. of (\ref{azbsigmaQgamma}) is preciely $a_{\bar{z}}\left[\hat{\gamma}{\tau};\ \frac{\mu_s}{ (c\tau+d)^s};\ (c\tau+d)^sQ_s \right]$, which proves the second identity of (\ref{transgamma}).

With (\ref{transgamma}) proven, evaluating the vectors of eigenvalues of both sides of (\ref{transgamma}) then gives the generalization to the pair (\ref{azrescale}) and (\ref{azbarrescale}):
\begin{equation}\label{rescalegamma}
\begin{aligned}
(c\tau+d)\  \Lambda\left(a_{z}\left[Q_s\right]\right) &=  \Lambda\left(a_{z}\left[(c\tau+d)^s Q_s\right]\right)\,,\\
(c\bar{\tau}+d)\  \Lambda \left(a_{\bar{z}}\left[\tau;\, \mu_s;\ Q_s\right]\right)&=\Lambda \left(a_{\bar{z}}\left[\hat{\gamma}{\tau};\ \frac{\mu_s}{ (c\tau+d)^s};\ (c\tau+d)^sQ_s \right]\right)\,.
\end{aligned}
\end{equation}

Having proved the pair (\ref{rescalegamma}), we can now use them to rewrite the holonomy condition of the solution $\gamma$ into the form of the conical surplus:
\begin{equation}\label{holoCSgamma}
\begin{aligned} 
i\ \vec{n}=\Lambda\left(\omega_{\textrm{A}}\left[ \tau;\, \mu_s;\ Q_s \right]\right)=\Lambda\left(\omega_{\phi}\left[\hat{\gamma}{\tau};\ \frac{\mu_s}{ (c\tau+d)^s};\ (c\tau+d)^sQ_s \right]\right)\,.
\end{aligned}
\end{equation}
Namely, the solution $\gamma$ can be generated by a passive change of variables (\ref{modularfull}) on the conical surplus.

In order to sum over the contributions from all members of the `$\textrm{SL}(2,\mathbb{Z})$' family to the full partition function, we need to place all the solutions in a common grand canonical ensemble, with common temperature and chemical potentials. This requires us to switch to the active viewpoint of the transformation, i.e. we hold $\{\vec{n};\, \tau;\,\mu_s\}$ fixed and ask how the conserved charges $Q_s$ transform. Comparing (\ref{holoCSgamma})  with the holonomy condition (\ref{defineCS2}) of the conical surplus we conclude that for given $\{\vec{n};\, \tau;\,\mu_s\}$ the on-shell value of the charges $Q_s$ of this new solution $\gamma$ is related to that of the conical surplus via:
\begin{equation}\label{modularcharge}
Q^{\textrm{CS}}_t= q_t\left[\vec{n};\ \tau; \ \mu_{s}\right] \qquad \Longleftrightarrow \qquad Q^{\gamma}_t=\frac{1}{(c\tau+d)^t} q_t\left[\vec{n};\ \hat{\gamma}\tau; \ \frac{\mu_{s}}{(c\tau+d)^s}\right]  \qquad t=2,\ldots,N
\end{equation}

To summarize, we have proved that starting with a conical surplus with parameters $\{\vec{n};\tau;\mu_s\}$ (holonomy vector around $\phi$-cycle, boundary modulus, chemical potentials), the transformation (\ref{gamma}) maps it into another smooth constant solution with the same chemical potentials $\{\mu_s\}$, whose holonomy around the new A-cycle $z \sim z + 2 \pi (c\tau+d)$ is trivial and is given by the same vector $\vec{n}$, and whose on-shell values of charges are given by
\begin{equation}\label{chargegamma}
\textrm{solution $\gamma$: }\qquad Q^{\gamma}_t=\frac{1}{(c\tau+d)^t} q_t\left[\vec{n};\ \hat{\gamma}\tau; \ \frac{\mu_{s}}{(c\tau+d)^s}\right]  \qquad t=2,\ldots,N \,.
\end{equation}
where $ q_t$ is the function defined by the charge function of the conical surplus as in (\ref{chargeCS}).

Then as $\gamma$ runs through $\Gamma_{\infty}\backslash\Gamma$, we obtain a `$\textrm{SL}(2,\mathbb{Z})$' family of smooth constant solutions with a common set of $\{\vec{n};\ \tau;\  \mu_s \}$; their only difference is in the choices of their A/B cycles which in turn give different on-shell values for the charges as given by (\ref{chargegamma}). Within each such `$\textrm{SL}(2,\mathbb{Z})$' family (labeled by $\{\vec{n};\tau;\mu_s\}$), all except for the one with $\gamma=\tiny\begin{pmatrix}
1 & 0\\
0& 1 
\end{pmatrix}$ (i.e. the conical surplus) are higher-spin black holes (since their A-cycles all have $t$-direction), in complete parallel to the spin-2 case.

\bigskip
\subsection{Reasoning from dual CFT}\label{sec:CFT}

Let us see this from the dual CFT side, borrowing the argument from  \cite{Dijkgraaf:1996iy}. In the CFT side, switching on the chemical potentials $\mu_s$ of the higher-spin charges can be accounted for perturbatively by adding an irrelevant perturbation to the  action
\begin{equation}
S_{CFT} \rightarrow S_{CFT}+ \sum^N_{s\geq 3}\int \frac{d^2z}{4 \pi \tau_2} \, \left( \mu_s W^{(s)}(z)- c.c.\right)\,.
\end{equation}
Correspondingly, the partition function (\ref{Zbulk}) is equal to the torus amplitude in the CFT side: 
\begin{equation}
Z^{\textrm{CFT}}[\tau;\,\mu_s]\equiv \langle e^{2 \pi i \left(\sum^N_{s= 3}\mu_s \int \frac{d^2z}{2 \pi \tau_2} W^{(s)}(z)- c.c.\right)}  \rangle 
\end{equation}

The volume element $\frac{d^2z}{2\pi \tau_2}$ should be invariant under a modular transformation; therefore the  modular transformation (\ref{taumodular}) induces a transformation of the coordinates:
\begin{equation}\label{zmodular}
\tau \longmapsto \hat{\gamma}\tau=\frac{a\tau+b}{c\tau+d}  \qquad \Longrightarrow  \qquad z \longmapsto z'=\frac{z}{c\tau+d}\,.
\end{equation}
In the highest-weight gauge we are using, the spin-$s$ field $W^{(s)}(z)$ is a Virasoro primary of  weight-$s$, therefore under (\ref{zmodular})
\begin{equation}
W^{(s)}(z)\longmapsto W^{(s)\prime}(z^{\prime})=(c\tau+d)^s W^{(s)}(z) \,.
\end{equation}
Namely $W^{(s)}(z)$ has weight-$s$ under both conformal and modular transformations \cite{Dijkgraaf:1996iy}. First this means that $Q_s=\int \frac{d^2z}{2 \pi \tau_2} W^{(s)}(z)$ 
also transforms as 
\begin{equation}\label{Qmap}
Q_s\longmapsto Q^{\prime}_s=(c\tau+d)^s Q_s\,.
\end{equation} Second the invariance of the integrand implies that the chemical potential $\mu_s$ transforms as
\begin{equation}\label{mumodular}
\mu_s \longmapsto \mu'_s=\frac{1}{(c\tau+d)^s}\mu_s\,.
\end{equation}
The maps (\ref{zmodular}), (\ref{Qmap}), and (\ref{mumodular}) together are precisely what we have shown earlier in (\ref{modularfull}) to be the required change of variables (in the passive viewpoint of the transformation) to map the conical surplus to the solution $\gamma$. 

Finally we make a comparison between the transformations in the passive viewpoint and the active one. In the passive viewpoint, one applies the following change of variables:
\begin{equation}\label{fieldredef}
\textrm{passive:}\qquad \vec{n}\,\,\textrm{fixed}\,, \quad \tau \longmapsto \hat{\gamma}\tau= \frac{a\tau+b}{c\tau+d}\,,\quad    \mu_s \longmapsto \frac{\mu_s}{(c\tau+d)^s}\,, \quad  Q_s \longmapsto (c\tau+d)^sQ_s \,.
\end{equation}
We have adopted the passive viewpoint in the proof of the mapping between different members of solutions under the modular transformation. Once the mapping is established, we switch to the active viewpoint in order to place all solutions in the `$\textrm{SL}(2,\mathbb{Z})$' family in one grand canonical ensemble:
\begin{equation}\label{bonafide}
\begin{aligned}
\textrm{active:}\qquad &\{\vec{n};\,\tau;\,\mu_s\}\quad \textrm{fixed}\,,\\
& Q_s=q_s\left[\vec{n};\,\tau;\, \mu_r\right] \,\longmapsto \, Q^{\gamma}_s=\frac{1}{(c\tau+d)^s}q_s\left[\vec{n};\, \hat{\gamma}{\tau};\ \frac{\mu_r}{ (c\tau+d)^r} \right]\,.
\end{aligned}
\end{equation}



\bigskip
\subsection{Coordinate transformations between members of `$\textrm{SL}(2,\mathbb{Z})$' family}\label{sec:gammacoord}

In Section~\ref{sec:Scoord} we proved that a black hole with parameter $\{\vec{n};\ \tau;\ \mu_s\}$  can be mapped to a conical surplus with  parameter $\{\vec{n};\, -\frac{1}{\tau};\ \frac{\mu_s}{\tau^s}\}$ via a coordinate transformation (see (\ref{diffeoS})). The generalization to the full `$\textrm{SL}(2,\mathbb{Z})$' family is immediate. 

First,  the gauge field components $(a_z,a_{\bar{z}})$ of a conical surplus and those of solution $\gamma$ are related via:
 \begin{equation}\label{aCSgamma}
 \begin{aligned}
 a^{\textrm{CS}}_z\left[\vec{n};\, \hat{\gamma}{\tau};\ \frac{\mu_s}{ (c\tau+d)^s} \right]&=
(c\tau+d) \cdot (c\tau+d) ^{L_0}\, a^{\gamma}_z\left[\vec{n};\ \tau;\ \mu_s\right]\,(c\tau+d) ^{-L_0}\,,\\
a^{\textrm{CS}}_{\bar{z}}\left[\vec{n};\, \hat{\gamma}{\tau};\ \frac{\mu_s}{ (c\tau+d)^s} \right]&=(c\bar{\tau} +d) \cdot(c\tau+d) ^{L_0} \, a^{\gamma}_{\bar{z}}\left[\vec{n};\ \tau;\ \mu_s\right]\,(c\tau+d) ^{-L_0} \,.
\end{aligned}
\end{equation}
Similar to the S-dual case, we apply a coordinate transformation to the  conical surplus:
\begin{equation}
\rho\longmapsto \rho^{\gamma}=\rho+\ln |c\tau+d|\,, \qquad z\longmapsto z^{\gamma}=\frac{z}{c\tau+d}\,, \qquad\qquad \bar{z}\longmapsto \bar{z}^{\gamma}=\frac{\bar{z}}{c\bar{\tau}+d}\,.
\end{equation}
With the gauge component $(a_z,a_{\bar{z}})$ kept fixed, the one-form $a$  and $A$ in this new coordinate are
\begin{equation}
a^{(z,\bar{z})^{\gamma}} \equiv a_{z}dz^{\gamma}+a_{\bar{z}}d\bar{z}^{\gamma}\,, \qquad A^{(\rho,z,\bar{z})^{\gamma}}\equiv  e^{-\rho^{\gamma} L_0}a^{(z,\bar{z})^{\gamma}} e^{\rho^{\gamma} L_0}+L_0 d\rho^{\gamma}\,;
\end{equation}
and similarly for $\bar{a}$ and $\bar{A}$.
The gauge one-form $\{a,\bar{a}\}$ of the solution $\gamma$ with parameter $\{\vec{n};\ \tau;\ \mu_s\}
$ in the original coordinate $(\rho,z,\bar{z})$ is related to that of the conical surplus with parameter $ \{\vec{n};\, \hat{\gamma}{\tau};\ \frac{\mu_s}{ (c\tau+d)^s}\} $ but in the new coordinate $(\rho,z,\bar{z})^{\gamma}$ via:
\begin{equation}\label{fullgammacoorda}
\begin{aligned}
a^{\textrm{CS}}\left[\vec{n};\, \hat{\gamma}{\tau};\ \frac{\mu_s}{ (c\tau+d)^s} \right]^{(z,\bar{z})^{\gamma}}&=(c\tau+d) ^{L_0}a^{\gamma}\left[\vec{n};\ \tau;\ \mu_s\right]^{(z,\bar{z})}(c\tau+d) ^{-L_0}\,,\\
\bar{a}^{\textrm{CS}}\left[\vec{n};\, \hat{\gamma}{\tau};\ \frac{\mu_s}{ (c\tau+d)^s} \right]^{(z,\bar{z})^{\gamma}}&=(c\bar{\tau}+d) ^{-L_0}\bar{a}^{\gamma}\left[\vec{n};\ \tau;\ \mu_s\right]^{(z,\bar{z})}(c\bar{\tau}+d) ^{L_0}\,.
\end{aligned}
\end{equation}
 The full gauge one-forms $(A,\bar{A})$ of the solution $\gamma$ with parameter $\{\vec{n};\ \tau;\ \mu_s\}
$ in the original coordinate $(\rho,z,\bar{z})$ is then identical to that of the conical surplus with parameter $ \{\vec{n};\, \hat{\gamma}{\tau};\ \frac{\mu_s}{ (c\tau+d)^s}\} $ but in the new coordinate $(\rho,z,\bar{z})^{\gamma}$ up to an overall constant gauge transformation:
\begin{equation}\label{fullgammacoord}
\begin{aligned}
\hat{h}_{\gamma}^{-1}\cdot A^{\textrm{CS}}\left[\vec{n};\, \hat{\gamma}{\tau};\ \frac{\mu_s}{ (c\tau+d)^s}  \right]^{(\rho,z,\bar{z})^{\gamma}}\cdot \hat{h}_{\gamma}&= A^{\gamma}\left[\vec{n};\ \tau;\ \mu_s\right]^{(\rho,z,\bar{z})}\,,\\
\hat{h}_{\gamma}^{-1}\cdot\bar{A}^{\textrm{CS}}\left[\vec{n};\, \hat{\gamma}{\tau};\ \frac{\mu_s}{ (c\tau+d)^s}  \right]^{(\rho,z,\bar{z})^{\gamma}}\cdot \hat{h}_{\gamma}&= \bar{A}^{\gamma}\left[\vec{n};\ \tau;\ \mu_s\right]^{(\rho,z,\bar{z})}\,,
\end{aligned}
\end{equation}
with 
\begin{equation}
\hat{h}_{\gamma}=\left(\frac{c\bar{\tau}+d}{c\tau+d}\right)^{\frac{L_0}{2}}\,.
\end{equation}
This means that all constant solutions in the SL$(2,\mathbb{Z})$ family are locally identical: they are all discrete quotients of the $\textrm{SL}(N,\mathbb{R})\otimes \textrm{SL}(N,\mathbb{R})$. Explicitly, for constant solutions, the connection $A$ in the gauge (\ref{gauge}) can be written as $A=g^{-1}dg $ with $g\equiv e^{za_z+\bar{z}a_{\bar{z}}} b\in \textrm{SL}(N,\mathbb{R})$ and $b=e^{\rho L_0}$ \cite{quotient}. Since for the solution $\gamma$, the space has a B-cycle $(z,\bar{z})\sim(z+2\pi(a\tau+b), \bar{z}+2\pi(a\bar{\tau}+b))$, $g^{\gamma}$ satisfies
\begin{equation}\label{quotienting}
g^{\gamma} \sim  \left( b\,\textrm{Hol}_{\textrm{B}}(A^{\gamma})\, b^{-1} \right) \cdot g^{\gamma}\,,
\end{equation}
where $\textrm{Hol}_{\textrm{B}}(A^{\gamma})$ is the holonomy of $A^{\gamma}$ around the B-cycle:
\begin{equation}\label{hologammaB}
\textrm{Hol}_{\textrm{B}}(A^{\gamma})=b^{-1} e^{2\pi\omega^{\gamma}_{\textrm{B}}} b 
\end{equation}
with 
\begin{equation}
\omega^{\gamma}_{\textrm{B}}
=(c\tau+d)^{-L_0}\cdot \left( \hat{\gamma}\tau \,a^{\textrm{CS}}_z\left[\vec{n};\, \hat{\gamma}{\tau};\ \frac{\mu_s}{ (c\tau+d)^s}\right] +\hat{\gamma}\bar{\tau}\, a^{\textrm{CS}}_{\bar{z}}\left[\vec{n};\, \hat{\gamma}{\tau};\ \frac{\mu_s}{ (c\tau+d)^s}  \right] \right) \cdot (c\tau+d)^{L_0}\,.
\end{equation}
using the definition (\ref{omegaB}) and the relation (\ref{aCSgamma}), and similarly for the $\bar{A}$ sector. 
Namely, the solution $\gamma$ is the quotient of $\textrm{SL}(N,\mathbb{R})$ by a matrix that is given by the holonomy of $A^{\gamma}$ around the B-cycle conjugated by $b$.\footnote{Note that although the holonomy $\textrm{Hol}_{\textrm{B}}(A^{\gamma})$ has an explicit $\rho$-dependence, its eigenvalues do not. In (\ref{quotienting}), the conjugation by $b$ serves to remove the $\rho$-dependence and extract the information on the eigenvalues of $\textrm{Hol}_{\textrm{B}}(A^{\gamma})$.}
For fixed $\{\tau,\mu_s\}$, the matrix $(b \, \textrm{Hol}_{\textrm{B}}(A^{\gamma})\,b^{-1})$  is a representation of a point in $\Gamma_{\infty}\backslash\Gamma$.

This is the analogue of the spin-$2$ story: in the spin-$2$ case, a solution $\gamma$ in the  `$\textrm{SL}(2,\mathbb{Z})$' family can be mapped by a coordinate transformation into an AdS$_3$ with modular parameter $\frac{a\tau+b}{c\tau+d}$ (instead of $\tau$). All these solutions are discrete quotients of the AdS$_3$ space by elements of SL$(2,\mathbb{Z})$ \cite{Maldacena:1998bw,Dijkgraaf:2000fq}. 

\bigskip
\subsection{Mapping of the free energy}


Recall that the free energy of a conical surplus and that of its S-dual black hole are related by (\ref{SmapF}). Now we generalize this mapping to the entire `$\textrm{SL}(2,\mathbb{Z})$' family. 
Different members of the `$\textrm{SL}(2,\mathbb{Z})$' family share the same holonomy vector $\vec{n}$, around their respective A-cycles. Therefore although we have a number of different expressions for the free energy, the one we should use to study the mapping under a modular transformation is the one written explicitly in terms of the holonomy matrices, and in a form universal across the `$\textrm{SL}(2,\mathbb{Z})$' family --- the expression  (\ref{freeenergygamma}) is ideal for this purpose.

Within a `$\textrm{SL}(2,\mathbb{Z})$' family, all members share the same $\{\vec{n};\ \tau; \,\mu_s\}$ but differ in their modular parameters $\gamma$ and hence their charges $Q^{\gamma}_s$, as given in (\ref{chargegamma}).
To obtain the mapping between the free energies of different members, we first need to rewrite (\ref{freeenergygamma}) in terms of only $\{\vec{n}; \ \tau; \ \mu_s\}$ and $\gamma$. Since we already know that the charges $Q^{\gamma}_s$ for different $\gamma$ are related via (\ref{chargegamma}), we will first rewrite (\ref{freeenergygamma}) in terms of $\{\vec{n}; \ \tau; \ \mu_s; \ Q^{\gamma}_s\}$ and $\gamma$. 

Using 
\begin{equation}
\begin{aligned}
\frac{1}{2}\Tr\left[\omega_{\textrm{A}}\omega_{\textrm{B}}\right]=&(a\tau+b)(c\tau+d) Q^{\gamma}_2 + \frac{i\textrm{Re}[(a\tau+b)(c\bar{\tau}+d)] }{2\tau_2}\sum^{N}_{s=3}s \mu_s Q^{\gamma}_s\\
&+ (a\bar{\tau}+b)(c\bar{\tau}+d)\frac{\Tr\left[(a_{\bar{z}})^2\right]}{2} \,,
\end{aligned}
\end{equation}
and
\begin{equation}
-\frac{1}{2}\vec{n}^2=\frac{1}{2}\Tr\left[ (\omega_{\textrm{A}})^2 \right] =(c\tau+d)^2 Q^{\gamma}_2 + \frac{i|c\tau+d|^2}{2\tau_2}\sum^{N}_{s=3}s \mu_s Q^{\gamma}_s+ \frac{(c\bar{\tau}+d)^2}{2}\Tr\left[(a_{\bar{z}})^2\right] \,,
\end{equation}
we arrive at an identity
\begin{equation}
\begin{aligned}
\frac{1}{2}\Tr\left[\omega_{\textrm{A}} \omega_{\textrm{B}}\right]=-\frac{\vec{n}^2}{2}\hat{\gamma}\bar{\tau}+
\frac{2 i \tau_2}{|c\tau+d|^2}(c\tau+d)^2 Q^{\gamma}_2-\sum^N_{s=3}\frac{s}{2} \ \mu_s Q^{\gamma}_s\,,
\end{aligned}
\end{equation}
which immediately allows us to rewrite the free energy (\ref{freeenergygamma}) in terms of $\{\vec{n};\ \tau;\ \mu_s;\ Q^{\gamma}_s\}$ and $\gamma$:
\begin{equation}\label{Fmodulargamma}
\begin{aligned}
-\beta F^{\gamma}
=2\pi i k\Big(
\frac{2 i \tau_2}{|c\tau+d|^2}\left[\frac{\vec{n}^2}{2}+(c\tau+d)^2 Q^{\gamma}_2+(c\bar{\tau}+d)^2\bar{Q}^{\gamma}_2\right]-\sum^N_{s=3}(s-1) (\mu_s Q^{\gamma}_s-\bar{\mu}_s\bar{Q}^{\gamma}_s)\Big)\,.
\end{aligned}
\end{equation}
Then recall that $Q^{\gamma}_s$ depends on $\{\vec{n}; \ \tau; \ \mu_s\}$ and $\gamma$ via (\ref{chargegamma}), 
we obtain the expression of the free energy in terms of $\{\vec{n}; \ \tau; \ \mu_s\}$ only:
\begin{equation}\label{Fcovariant}
\begin{aligned}
&-\beta F^{\gamma}
=\, 2\pi i k\Big[\frac{2 i \tau_2}{|c\tau+d|^2}\left(\frac{\vec{n}^2}{2}+ q_2\left[\vec{n};\ \hat{\gamma}\tau;\ \frac{\mu_s}{(c\tau+d)^s}\right] 
+\bar{ q}_2\left[\vec{n};\ \hat{\gamma}\bar{\tau};\ \frac{\bar{\mu}_s}{(c\bar{\tau}+d)^s}\right]\right) 
\\
&\,\, -\sum^N_{s=3}(s-1) \left(\frac{\mu_s}{(c\tau+d)^s} \, q_s\left[\vec{n};\ \hat{\gamma}\tau;\ \frac{\mu_s}{(c\tau+d)^s}\right] -\frac{\bar{\mu}_s}{(c\bar{\tau}+d)^s}\,\bar{ q}_s\left[\vec{n};\ \hat{\gamma}\bar{\tau};\ \frac{\bar{\mu}_s}{(c\bar{\tau}+d)^s}\right]\right) \Big]\,.
\end{aligned}
\end{equation}
Comparing this with (\ref{FmodularCSGC}), we obtain the map between the free energy of the conical surplus and that of the solution $\gamma$;
\begin{equation}\label{modularF}
F^{\textrm{CS}}\left[\vec{n};\ \tau; \ \mu_{s}\right] = \mathcal{F}\left[\vec{n};\ \tau; \ \mu_{s}\right] \qquad \Longleftrightarrow \qquad F^{\gamma}\left[\vec{n};\ \tau; \ \mu_{s}\right] =\mathcal{F}\left[\vec{n};\ \hat{\gamma}{\tau};\ \frac{\mu_s}{(c\tau+d)^s}\right]\,.
\end{equation}	
This proves that different solutions $\gamma$ in the `$\textrm{SL}(2,\mathbb{Z})$' family share the same form for their free energies, and they can all be obtained by applying the following transformation 
\begin{equation}\label{fullmodulartaumu}
\textrm{modular transformation in GCE: }\qquad\tau \longmapsto \hat{\gamma}{\tau}=\frac{a\tau+b}{c\tau+d} \qquad \quad \mu_s \longmapsto \frac{\mu_s}{(c\tau+d)^s}\,,
\end{equation}
on the free energy of the conical surplus.

Recall that in the spin-$2$ case, the metric of the solution labeled by $\gamma$ in the `$\textrm{SL}(2,\mathbb{Z})$' family can be brought into the metric of the AdS$_3$ with modular parameter $\hat{\gamma}{\tau}=\frac{a\tau+b}{c\tau+d}$ (instead of $\tau$) via a coordinate transformation. Since the on-shell action should be invariant under the coordinate transformation, the free energies of the solution $\gamma$  and the AdS$_3$ is necessarily related by the modular transformation: $F^{\gamma}[\tau]=F^{\textrm{AdS}_3}[\frac{a\tau+b}{c\tau+d}]$, and in particular $F^{\textrm{BTZ}}[\tau]=F^{\textrm{AdS}_3}[-\frac{1}{\tau}]$.

The non-trivial part of the story is in the construction of the full action (including boundary terms) whose Euclidean on-shell action (hence the free energy) is invariant under coordinate transformations. Now we have seen a complete parallel in the $\mathfrak{sl}(N)$ Chern-Simons theory. As shown in Section~\ref{sec:gammacoord}, different members of the `$\textrm{SL}(2,\mathbb{Z})$' family are related by the coordinate transformation (\ref{fullgammacoord}), therefore the mapping between their free energies (\ref{modularF}) not only confirms the mapping between different solutions via (\ref{bonafide}) and (\ref{fullgammacoord}), more importantly, it provides strong evidence that the thermodynamics that we derived in Section~\ref{sec:thermodynamics} is consistent and applies to all members in the `$\textrm{SL}(2,\mathbb{Z})$' family.  We regard this as a much more non-trivial result than the mapping of different members at the level of solution.

\bigskip
\subsection{Modular invariant full partition function}

The full partition function should include contributions from all saddle points. Let us first classify all classical solutions known to us. First, for each holonomy vector $\vec{n}$ satisfying (\ref{ndef}) there is a `$\textrm{SL}(2,\mathbb{Z})$' family of solutions constructed earlier in this section. 
Using $Z=e^{-\beta F}$ and (\ref{modularF}), we see that the contribution from each member $\gamma$ is an image of the modular transformation  (\ref{fullmodulartaumu}) of the contribution from the conical surplus:
\begin{equation}
Z^{\gamma}_{\vec{n}}\left[ \tau; \ \mu_{s}\right] =Z^{\textrm{CS}}_{\vec{n}}\left[\hat{\gamma}{\tau};\ \frac{\mu_s}{(c\tau+d)^s}\right]\,.
\end{equation}
For given $\vec{n}$, we should first sum over all members within this `$\textrm{SL}(2,\mathbb{Z})$' family, which consists of all the modular images of the conical surplus: 
\begin{equation}
Z_{\vec{n}}\left[ \tau; \ \mu_{s}\right]=\sum_{\gamma \in \Gamma_{\infty}\backslash \Gamma} Z^{\gamma}_{\vec{n}}\left[ \tau; \ \mu_{s}\right]=\sum_{\gamma \in \Gamma_{\infty}\backslash \Gamma}Z^{\textrm{CS}}_{\vec{n}}\left[\hat{\gamma}{\tau};\ \frac{\mu_s}{(c\tau+d)^s}\right]\,.
\end{equation}
So far this is in complete parallel to the spin-$2$ case studied in \cite{Witten:2007kt,Maloney:2007ud}. 
Now comes the major difference: we should also sum over all distinct $\vec{n}$ satisfying (\ref{ndef}):
\begin{eqnarray}
Z\left[\tau;\, \mu_s\right]=\sum_{\vec{n}}Z_{\vec{n}}\left[ \tau; \ \mu_{s}\right]\,.
\end{eqnarray}
This expression is manifestly modular invariant. After the appropriate regularization, one can then use it to extract the phase structure of the full theory. For instance, the (infinite) sum over $\vec{n}$ might smear out the Hawking-Page transition, as argued in \cite{Banerjee:2012aj}.
We leave this to future work.

\bigskip
\section{Discussion}\label{sec:conclude}

The goal of this paper is simple: in the three-dimensional asymptotically AdS$_3$ space, we want to generalize the construction of the  `$\textrm{SL}(2,\mathbb{Z})$' family of solutions 
from the spin-$2$ gravity to the higher-spin gravity. We achieved this in the $\mathfrak{sl}(N)$ Chern-Simons theory. The main results have already been summarized in Section~\ref{sec:intro}, now we end with a discussion on the main difference of the higher-spin theory from the spin-$2$ one. 

In the spin-$2$ theory one usually works with the metric, and under a modular transformation of the boundary torus the mapping between different solutions can be written in terms of the coordinate transformations of the bulk metric. In the higher-spin theory, the metric (more precisely the line element $ds^2$) is no longer a gauge invariant concept therefore would not be suitable for a discussion on the modular property. However, as we have seen, the gauge connection in the Chern-Simon higher-spin theory
is actually more convenient for the study of the modular properties than the metric in the spin-$2$ gravity. 

First of all, the defining equation of a smooth solution (the trivial holonomy condition around its A-cycle) is manifestly modular covariant, i.e. a passive change of variables (\ref{modularfull}) directly maps it into the defining equation of another smooth solution, as shown in  (\ref{holoCSgamma}). Moreover, the free energy also has an expression (\ref{freeenergygamma}) that is universal for all solutions and using which one can arrive at a modular-covariant expression (\ref{Fcovariant}). For the entropy, we have also derived an expression (\ref{entropy}) that applies to all members of `$\textrm{SL}(2,\mathbb{Z})$' family.

Finally, we finish with a proof of the modular invariance of the integrability condition of the theory, and with it a discussion on an important difference between the `canonical' formalism and the `holomorphic' one.


\bigskip
\subsection{Modular invariance of integrability condition}

The existence of the free energy formula (\ref{freeenergygamma1})  guarantees the following integrability condition
\begin{equation}\label{integrable}
\frac{\partial Q_s}{\partial \mu_t}=\frac{\partial Q_t}{\partial \mu_s}\,, \qquad \qquad \frac{\partial T}{\partial \mu_s}=\frac{\partial Q_s}{\partial \tau}\,, \qquad s,t=3,\ldots,N\,,
\end{equation}
for all members of the `$\textrm{SL}(2,\mathbb{Z})$' family, i.e. the relations (\ref{integrable}) should be modular invariant. However, it is instructive to see how it works in detail. This not only serves as a non-trivial consistency check for our construction of the `$\textrm{SL}(2,\mathbb{Z})$' family of solutions; it will also be used later when we compare with other discussions of the thermodynamics in this theory. 


To show that (\ref{integrable}) is modular invariant, we compare the relation (\ref{integrable}) for a conical surplus with that of a generic solution labeled by $\gamma$,  and show that the latter can be obtained by a modular transformation (\ref{fullmodulartaumu}) of the former. First, let's start with the first condition of (\ref{integrable}). When the solution is a conical surplus, we plug (\ref{chargeCS}) into (\ref{integrable}) and obtain  
\begin{equation}\label{IC1CS}
\frac{\partial q_s}{\partial \mu_t}[\tau, \mu_r]=\frac{\partial q_t}{\partial \mu_s}[\tau,\mu_r]\,.
\end{equation}
(In this proof we omit $\vec{n}$ since it stays invariant under modular transformations.) Then for a generic solution labeled by $\gamma$, plugging in (\ref{chargegamma}) and using\footnote{A clarification of notation: the derivative always takes place before the change of variables, e.g. $\frac{\partial q_s}{\partial \mu_t}[\hat{\gamma}{\tau}; \ \frac{\mu_r}{(c\tau+d)^r}]\equiv \frac{\partial q_s}{\partial \mu_t}[\tau; \ \mu_r]\vert_{\tau \mapsto\hat{\gamma}{\tau}; \ \mu_r \mapsto \frac{\mu_r}{(c\tau+d)^r}}$.}
\begin{equation}\label{dQgamma}
\frac{\partial Q^{\gamma}_s}{\partial \mu_t}=\frac{1}{(c\tau+d)^{s+t}}\frac{\partial q_s}{\partial \mu_t}[\hat{\gamma}{\tau}; \ \frac{\mu_r}{(c\tau+d)^r}]\,, 
\end{equation}
we get the integrability condition to be
\begin{equation}\label{IC1gamma}
\frac{\partial q_s}{\partial \mu_t}[\hat{\gamma}{\tau}; \ \frac{\mu_r}{(c\tau+d)^r}] = \frac{\partial q_t}{\partial \mu_s}[\hat{\gamma}{\tau}; \ \frac{\mu_r}{(c\tau+d)^r}]\,,\end{equation}
which is precisely the modular transformation (\ref{fullmodulartaumu}) of the integrability condition of the conical surplus (\ref{IC1CS}).

The second condition of (\ref{integrable}) is slightly harder since it involves $T$, which is not modular covariant in itself: for a generic solution labeled by $\gamma$, $T$ is
\begin{equation}
T=\frac{1}{(c\tau+d)^2}\left[\frac{\vec{n}^2}{2}+ q_2+\bar{ q}_2\right]+\frac{i}{2 \tau_2}\sum^N_{s=3}s\left[\frac{\mu_s q_s}{(c\tau+d)^s}-\frac{c\bar{\tau}+d}{c\tau+d}\frac{\bar{\mu}_s\bar{ q}_s}{(c\bar{\tau}+d)^s}\right]\,,
\end{equation}
where the variables inside the function $ q_t$ is $(\hat{\gamma}\tau; \, \frac{\mu_r}{(c\tau+d)^r})$.  However as we will now show, the integrability condition involving $T$ is nevertheless modular invariant. First, for the conical surplus 
\begin{equation}\label{dTCS}
\frac{\partial T}{\partial \mu_s}=\frac{\partial  q_2}{\partial \mu_s}[\tau, \mu_r]+\frac{i}{2\tau_2}\mathcal{G}_s[\tau, \mu_r]\,,
\end{equation}
with $\mathcal{G}_s$ defined as
\begin{equation}
\mathcal{G}_s[\tau, \mu_r]\equiv s  q_s[\tau, \mu_r]+\sum^N_{t=3}\ t\ \mu_t \ \frac{\partial  q_t}{\partial \mu_s}[\tau, \mu_r]\,,
\end{equation}
and the integrability condition is
\begin{equation}\label{IC2CS}
\left(\frac{\partial  q_2}{\partial \mu_s}[\tau, \mu_r]+\frac{i}{2\tau_2}\mathcal{G}_s[\tau, \mu_r]\right)-\frac{\partial  q_s}{\partial \tau}[\tau, \mu_r]=0\,.
\end{equation}

Now for a generic solution, $\frac{\partial T}{\partial \mu_s}$ becomes
\begin{equation}\label{dTgamma}
\begin{aligned}
\frac{\partial T}{\partial \mu_s}&=\frac{1}{(c\tau+d)^{2+s}}\left(\frac{\partial  q_2}{\partial \mu_s}[\hat{\gamma}{\tau}; \ \frac{\mu_r}{(c\tau+d)^r}]+\frac{i(c\tau+d)^2}{2\tau_2}\mathcal{G}_s[\hat{\gamma}{\tau}; \ \frac{\mu_r}{(c\tau+d)^r}]\right)\,.
\end{aligned}
\end{equation}
Comparing (\ref{dTgamma}) with (\ref{dTCS}) shows that $\frac{\partial T}{\partial \mu_s}$ is almost covariant, but not quite: the factor in front of $\mathcal{G}_s$ is $\frac{i(c\tau+d)^2}{2\tau_2}$ instead of $\frac{i|c\tau+d|^2}{2\tau_2}$. On the other hand, $\frac{\partial Q^{\gamma}_s}{\partial \tau}$ becomes 
\begin{equation}\label{dQs}
\begin{aligned}
\frac{\partial Q^{\gamma}_s}{\partial \tau}&=\frac{1}{(c\tau+d)^{2+s}}\left(\frac{\partial  q_s}{\partial \tau}[\hat{\gamma}{\tau}; \ \frac{\mu_r}{(c\tau+d)^r}]-c(c\tau+d)\mathcal{G}_s[\hat{\gamma}{\tau}; \ \frac{\mu_r}{(c\tau+d)^r}]\right)\,.
\end{aligned}
\end{equation}
In general the derivative (w.r.t. $\tau$) of a modular form is not a modular form. We see something similar happens here as well: although $Q_s$ transforms covariantly under the modular transformation, the last terms of the above equation (\ref{dQs}) says that $\frac{\partial Q^{\gamma}_s}{\partial \tau}$ does not. These two oddities cancel each other out when we subtract (\ref{dQs}) from (\ref{dTgamma}) to write down the integrability condition for the solution $\gamma$:
\begin{equation}\label{IC2CStrans}
0=\frac{\partial  q_2}{\partial \mu_s}[\hat{\gamma}{\tau}; \ \frac{\mu_r}{(c\tau+d)^r}]+\frac{i|c\tau+d|^2}{2\tau_2}\mathcal{G}_s[\hat{\gamma}{\tau}; \ \frac{\mu_r}{(c\tau+d)^r}]-\frac{\partial  q_s}{\partial \tau}[\hat{\gamma}{\tau}; \ \frac{\mu_r}{(c\tau+d)^r}]\,,\end{equation}
which is precisely the modular transformation (\ref{fullmodulartaumu}) of the integrability condition of the conical surplus (\ref{IC2CS}).

\subsection{Canonical vs. holomorphic}
Now we would like to compare the `holomorphic' formalism with the `canonical' one (which we used throughout this paper). The main difference between these two approaches is in the identification of the spin-$2$ conserved charges: $Q_2$ in the `holomorphic' approach and $T$ (as defined in (\ref{TTbardef})) in the `canonical' one. This difference in turn leads to different results for the entropy and the free energy.
Now relevant to this paper we will focus on the difference in the modular properties of these two formalisms. 


The first tell-tale sign that the modular transformation (\ref{fullmodulartaumu}) would not be applicable in the `holomorphic' formalism is from the integrability condition. It suffices to look at the $\mathfrak{sl}(3)$ theory considered in \cite{Gutperle:2011kf}. For the black hole in this theory, the spin-$2$ conserved charge is the holomorphic $Q_2$, which satisfies the integrality condition \cite{Gutperle:2011kf}
\begin{equation}\label{inteholo}
\frac{\partial Q_2}{\partial \mu_3} =  \frac{\partial Q_3}{\partial \tau}\,.
\end{equation}
However, if $Q_s$ was to transform under the modular transformation (\ref{fullmodulartaumu}) in the same way as what we proposed:
\begin{equation}
Q_s=q_s[\tau;\, \mu_r] \quad \longmapsto \quad Q^{\gamma}_s=\frac{1}{(c\tau+d)^s}q_s[\hat{\gamma}\tau; \, \frac{\mu_r}{(c\tau+d)^r}]
\end{equation}
then (\ref{inteholo}) would not be modular invariant --- the l.h.s. of (\ref{inteholo}) is modular covariant as shown in (\ref{dQgamma}) whereas the r.h.s. is not, as explained in (\ref{dQs}) and the lines below. 
This suggests that if we are to extend the result of \cite{Gutperle:2011kf} into a full `$\textrm{SL}(2,\mathbb{Z})$' family, the modular transformation would not be the same as given in the present paper. It would be interesting to work out the appropriate modular transformation in the `holomorphic' formalism, for which the translation between the two formalisms discussed in \cite{deBoer:2013gz} would be helpful.\footnote{Note added: the authors in \cite{Compere:2013nba} proposed another canonical formalism in which the charges sit in $a_{\phi}$ and the chemical potentials in $\omega_{t}$. It would be interesting to try and construct a modular family in this formalism. }

\bigskip
\section*{Acknowledgements}

We take great pleasure in thanking Matthias Gaberdiel and Stefan Theisen for very helpful discussions. WL also thanks Galileo Galilei Institute for Theoretical Physics and the Benasque string workshop for their hospitality during various stages of this work. FLL is supported by Taiwan's NSC grants (grant NO. 100-2811-M-003-011).
\bigskip

\appendix

\section{Solving $\sigma_s$ in terms of $\mu_s$}
\label{app:sigmamu}
An identity that is central to the proof of this paper is eq. (\ref{muQtosigma}). 
To prove it, we can equivalently prove its inverse:
\begin{equation}\label{sigmatomu}
\frac{i}{2\tau_2}\mu_s =
\sum^N_{s^{\prime}=s}  \sigma_{s^{\prime}} \ \tilde{H}_{s^{\prime}-s} (Q_t)
\end{equation}
where $\tilde{H}_{s^{\prime}-s} (Q_t)$ is another homogenous polynomial of degree-$(s^{\prime}-s)$ with variables $Q_t$ having degree-$t$.

First recall that the $N-1$ equations (\ref{Nminus1}) with $a_{\bar{z}}$ given in  (\ref{abarz}) determine $\sigma_s$ in terms of $\mu_s$. Plugging (\ref{abarz}) into (\ref{Nminus1}), and using $\Tr[W^{(s)}_m]=0$ and $a_{z}=L_1+\mathbf{Q}$, we have
\begin{equation}\label{trace}
\frac{1}{t^{(s)}}\sum^N_{s^{\prime}=2}\sigma_{s^{\prime}}\Tr\left[W^{(s)}_{-s+1} \left(L_1+\mathbf{Q}\right)^{s^{\prime}-1} \right]=\frac{i}{2 \tau_2} \mu_s\qquad s=2,\ldots,N
\end{equation}
where $\mu_2\equiv 0$. Then plugging in $\mathbf{Q}=\sum^N_{s=2} \frac{Q_s}{t^{(s)}} W^{(s)}_{-s+1}$ into above and using the fact the trace only picks up the zero modes, namely
\begin{equation}
\Tr\left[W^{(s)}_{-s+1} \left(L_1+\mathbf{Q}\right)^{s^{\prime}-1} \right] =\sum^{s'-1}_{m=0}\ c_{s',s}\cdot(\prod^{m}_{j=1}Q_{s_j})\cdot \delta_{-s+1+(s'-1-m)+\sum^{m}_{j=1}(-s_j+1),\,0}
\end{equation}
where $c_{s,s^{\prime}}$ are some rational numbers which can be computed from (\ref{Wexplicit}) but whose explicit values do not concern us here. Therefore the trace in (\ref{trace}) contains only terms  $\prod^{r}_{m=1}Q_{s_j}$ with $\sum^{m}_{j=1}s_j=s^{\prime}-s$, i.e. it is a homogenous polynomial of degree-$(s^{\prime}-s)$ with variables $Q_t$ having degree-$t$. Thus we have proved (\ref{sigmatomu}), which in turns gives (\ref{muQtosigma}).

\bigskip
\section{$\textrm{SL}(2,\mathbb{Z})$ family in spin-$2$ case}\label{app:spin2}

Now we review how all BTZ black holes in spin-2 gravity can be obtained from modular transformations of the AdS$_3$ space. Below mostly follows the exposition of \cite{Dijkgraaf:2000fq}.

The thermal AdS$_3$ is a solid torus, with metric 
\begin{equation}\label{cosetmetric}
ds^2
=d \rho^2+ du^2+d \bar{u}^2+2 du d\bar{u} \cosh \rho  
\end{equation}
where $u=\frac{i}{2}(\phi+i t_E)$ with $\phi$ the angular coordinate and $t_E$ the Euclidean time. In terms of $u$ the contractible and non-contractible cycles are
\begin{eqnarray}\label{ABcycleu}
\textrm{Contractible(A)-cycle: }&& \qquad 2u\sim 2u + 2 \pi i  \\
\textrm{Non-contractible(B)-cycle: } &&\qquad 2u \sim 2u + 2 \pi i s 
\end{eqnarray}
where $s=s_1+is_2$ (with $s_2=2\pi\beta\geq 0$) is the modulus of the boundary torus in this homology basis (contractible cycle, non-contractible cycle).

Starting with the thermal AdS$_3$ solution, all asymptotically AdS$_s$ solutions (including AdS$_3$ and all BTZ black holes) can be obtained via modular transformations. The easiest way to see this is the following. Since all asymptotically AdS$_s$ solutions are locally diffeomorphic, they share the same metric (\ref{cosetmetric}) but with different maps of $u$ to $z\equiv \phi+it_E$.
Since $s$ can be mapped to a unique point $\tau$ in the fundamental domain via 
\begin{equation}
s=\frac{a\tau+b}{c\tau+d} \qquad \qquad  \begin{pmatrix}a& b\\c &d\end{pmatrix}\in \textrm{PSL}(2,\mathbb{Z})
\end{equation} 
we can uniquely define
\begin{equation}\label{ztou}
2u= \frac{i}{c\tau+d} z \qquad \qquad z=\phi + i t_E
\end{equation}
First, the A/B cycle in terms of $u$ translate into the A/B cycle in terms of $z$: 
\begin{equation}
\begin{aligned}
\textrm{Contractible(A)-cycle: }&& \qquad  z\sim z + 2 \pi (c \tau +d)  \\
\textrm{Non-contractible(B)-cycle: } &&\qquad z \sim z + 2 \pi  (a \tau+b)
\end{aligned}
\end{equation}
Namely, the map plus the designation of A/B-cycles (\ref{ABcycle}) determines which space-time cycle is non-contractible ($\phi$, or $t$, or a combination of the two), thus tell us whether the geometry is a thermal AdS$_3$ or a BTZ black hole. 

This can be confirmed by directly computing the metric. The metric in terms of $(\rho, t_E,\phi)$, obtained by plugging (\ref{ztou}) into (\ref{cosetmetric}), can be brought into the BTZ form via another local coordinate transformation (see \cite{Dijkgraaf:2000fq} for details):
\begin{eqnarray}\label{BTZform}
&&ds^2=N^2(r) dt^2_{E}+\frac{dr^2}{N^2(r)}+r^2 (d\phi + N^{\phi}(r)dt_{E})^2 \\
&&N^2(r)=\frac{(r^2-r^2_{2})(r^2+r^2_{1})}{r^2}, \qquad N^{\phi}(r)=\frac{r_1r_2}{r^2},\qquad\textrm{with } \,r_1+ir_2=\pm \frac{1}{c\tau+d}\nonumber
\end{eqnarray}
where $(c,d)=(0,1)$ corresponds to the thermal AdS$_3$ and $(c,d)=(1,0)$ to the BTZ black hole with $z\sim z+ 2 \pi (-\frac{1}{\tau})$; and other $(c,d)$ with $\textrm{gcd}(c,d)=1$ gives the whole ``$\textrm{SL}(2,\mathbb{Z})$" family of AdS$_3$ and BTZ black holes.

Once we write down the AdS$_3$ solution, we can generate the entire family via modular transformation.
The full partition function is the sum over all modular images:
\begin{equation}
Z
=\sum_{\Gamma_{\infty}\backslash \Gamma} Z_{\textrm{AdS}_3}(\frac{a \tau+b}{c \tau +d})
\end{equation}
The resulting partition function is divergent and need to be regularized. We refer this issue to \cite{Dijkgraaf:2000fq, Manschot:2007ha}.

\bigskip

\section{Example-2: $N=4$}\label{app:sl4}

Now let's check the next simplest example: the $N=4$ case. The computation is essentially the same as the previous $N=3$ case, only with longer expressions.
First, $a_z$ is
\begin{equation}
a_z = L_1 - \frac{Q_2}{10}L_{-1} + \frac{Q_3}{24} W^{(3)}_{-2} - \frac{Q_4}{36} W^{(4)}_{-3}\,,
\end{equation}
And $a_{\bar{z}}$ is solved via (\ref{Nminus1}):
\begin{equation}
a_{\bar{z}} = \frac{i}{2 \tau_2}\Big[ -\frac{41}{50} \mu_4 Q_2\  a_z +\mu_3(a_z^2-\frac{\Tr\left[(a_z)^2\right]}{4}\mathbf{1} )+\mu_4(a_z^3-\frac{\Tr\left[(a_z)^3\right]}{4}\mathbf{1} )\Big]\,,
\end{equation}
Note the appearance of linear term of $a_{z}$, which is absent in the $a_{\bar{z}}$ for $N=3$ case (\ref{azbarN3}).

Now let's repeat the procedure for $N=3$ case. First the holonomy vector is now
\begin{equation}
\vec{n}=(n_2,n_1,-n_1,-n_2) \qquad  \textrm{with }\qquad n_i \in \mathbb{N}\, , \quad n_2 > n_1\,.
\end{equation}
For the conical surplus, the holonomy condition along the $\phi$-cycle ((\ref{holphi}) and (\ref{omegaphidef})) is equivalent to the following three equations
\begin{equation}\label{hol_sl4CS}
\Tr\left[(a_z + a_{\bar{z}})^2\right] = -2 \sum^2_{i=1}n_i^2\,, \quad \Tr\left[(a_z + a_{\bar{z}})^3\right] =0 \ ,\quad\Tr\left[(a_z + a_{\bar{z}})^4\right] = 2 \sum^2_{i=1}n_i^4
\end{equation}
which determines the charges $\{Q_2,Q_3,Q_4\}$ in terms of $\{\mu_3,\mu_4 \}$ and $\tau$. 
We will omit the rather long expressions for this result but simply plug them into the free energy (\ref{FmodularCS}) to get the final answer for the free energy (in terms of a power expansion of $\{\mu_3,\mu_4\}$):
\begin{equation}\label{fcs_sl4}
\begin{aligned}
-\beta F^{\textrm{CS}} & = 4 \pi k \cdot \tau_2\cdot\left[ (n_1^2+n_2^2) -\frac{(n_1^2-n_2^2)^2}{2} (\alpha_3^2+\bar{\alpha}_3^2) + (n_1^2-n_2^2)^2(n_1^2+n_2^2)\,(\alpha_3^4+\bar{\alpha}_3^4)  \right.  \\
& ~+~ \frac{(9 n_1^4 - 82  n_1^2 n_2^2 + 9  n_2^4)}{100}\,(\alpha_4 - \bar{\alpha}_4) - \frac{7\,(n_1^2-n_2^2)^2 (n_1^2+n_2^2)}{10}\,(\alpha_3^2\, \alpha_4-\bar{\alpha_3}^2\, \bar{\alpha}_4)   \\
& ~+~ \frac{(n_1^2+n_2^2)(81\, n_1^4 + 862\, n_1^2 n_2^2 + 81\, n_2^4)}{2500}\,(\alpha_4^2+\bar{\alpha}_4^2)  \\
&  \left. ~-~ \frac{(n_1^2-n_2^2)^2 (89\,n_1^4 -2\,n_1^2 n_2^2 + 89\, n_2^4)}{100}\,(\alpha_3^2 \alpha_4^2 + \bar{\alpha}_3^2\bar{\alpha}_4^2) + \cdots \right] \,.
\end{aligned}
\end{equation}
with
\begin{equation}
\alpha_{3}\equiv\frac{i}{2\tau_2}\mu_3 \qquad \qquad \alpha_{4}\equiv\frac{i}{2\tau_2}\mu_4
\end{equation}

Now we turn to the black hole. The conditions that solves $\{Q_2,Q_3\}$ is now the holonomy condition around the cycle $z \sim z+ 2\pi \tau$:
\begin{equation}\label{hol_sl4BH}
\Tr\left[(\tau a_z + \bar{\tau}a_{\bar{z}})^2\right] = -2 \sum^2_{i=1}n_i^2\,, \quad \Tr\left[(\tau a_z + \bar{\tau} a_{\bar{z}})^3\right] =0 \ ,\quad\Tr\left[(\tau a_z + \bar{\tau}a_{\bar{z}})^4\right] = 2 \sum^2_{i=1}n_i^4
\end{equation}
Again we will only write the final answer of the free energy:
\begin{equation}\label{fbh_sl4}
\begin{aligned}
-\beta F^{\textrm{BH}} & =  4 \pi k \cdot \frac{\tau_2}{|\tau|^2}\cdot \left[  (n_1^2+n_2^2) -\frac{(n_1^2-n_2^2)^2}{2} (\beta_3^2+\bar{\beta}_3^2) + (n_1^2-n_2^2)^2(n_1^2+n_2^2)\,(\beta_3^4+\bar{\beta}_3^4)  \right.  \\
& ~+~ \frac{(9 n_1^4 - 82  n_1^2 n_2^2 + 9  n_2^4)}{100}\,(\beta_4 - \bar{\beta}_4) - \frac{7\,(n_1^2-n_2^2)^2 (n_1^2+n_2^2)}{10}\,(\beta_3^2\, \beta_4-\bar{\beta_3}^2\, \bar{\beta}_4)   \\
& ~+~ \frac{(n_1^2+n_2^2)(81\, n_1^4 + 862\, n_1^2 n_2^2 + 81\, n_2^4)}{2500}\,(\beta_4^2+\bar{\beta}_4^2)  \\
&  \left. ~-~ \frac{(n_1^2-n_2^2)^2 (89\,n_1^4 -2\,n_1^2 n_2^2 + 89\, n_2^4)}{100}\,(\beta_3^2 \beta_4^2 + \bar{\beta}_3^2\bar{\beta}_4^2) + \cdots \right] \,.
\end{aligned}
\end{equation}
with
\begin{equation}
\beta_{3}\equiv\frac{i|\tau|^2}{2\tau_2}\frac{\mu_3}{\tau^3} \qquad \qquad \beta_{4}\equiv\frac{i|\tau|^2}{2\tau_2}\frac{\mu_4}{\tau^4}
\end{equation}
We see the free energy of the black hole is exactly the S-transformation
\begin{equation}\label{st_sl4}
\tau \longmapsto - \frac{1}{\tau}\,,\qquad \mu_3 \longmapsto  \frac{\mu_3}{\tau^3}\,  ,\qquad
\mu_4  \longmapsto  \frac{\mu_4}{\tau^4} \,.
\end{equation}
of the conical surplus answer (\ref{st_sl4}).

\begin{singlespace}

\end{singlespace}


\begin{thebibliography}{99}





\bibitem{Cangemi:1992my}
D.~Cangemi, M.~Leblanc and R.~B.~Mann,
``Gauge Formulation of the Spinning Black Hole in (2+1)-Dimensional Anti-de~Sitter Space,''
Phys.\ Rev.\ D {\bf 48} (1993) 3606
[gr-qc/9211013].


\bibitem{Banados:1992gq}
M.~Ba\~nados, M.~Henneaux, C.~Teitelboim and J.~Zanelli,
``Geometry of the (2+1) Black Hole,''
Phys.\ Rev.\ D {\bf 48} (1993) 1506
[gr-qc/9302012].

\bibitem{Carlip:1994gc}
S.~Carlip and C.~Teitelboim,
``Aspects of Black Hole Quantum Mechanics and Thermodynamics in (2+1)-Dimensions,''
Phys.\ Rev.\ D {\bf 51} (1995) 622
[gr-qc/9405070].

\bibitem{Steif:1995zm}
A.~R.~Steif,
``Supergeometry of Three-Dimensional Black Holes,''
Phys.\ Rev.\ D {\bf 53} (1996) 5521
[hep-th/9504012].


\bibitem{Maldacena:1998bw} 
  J.~M.~Maldacena and A.~Strominger,
  ``AdS(3) black holes and a stringy exclusion principle,''
  JHEP {\bf 9812}, 005 (1998)
  [hep-th/9804085].
  

\bibitem{Dijkgraaf:2000fq}
R.~Dijkgraaf, J.~M.~Maldacena, G.~W.~Moore and E.~P.~Verlinde,
``A Black Hole Farey Tail,''
hep-th/0005003.

\bibitem{Hawking:1982dh} 
  S.~W.~Hawking and D.~N.~Page,
  ``Thermodynamics of Black Holes in anti-De Sitter Space,''
  Commun.\ Math.\ Phys.\  {\bf 87}, 577 (1983).



\bibitem{Maloney:2007ud} 
  A.~Maloney and E.~Witten,
  ``Quantum Gravity Partition Functions in Three Dimensions,''
  JHEP {\bf 1002}, 029 (2010)
  [arXiv:0712.0155 [hep-th]].









\bibitem{Vasiliev:1989re} 
  M.~A.~Vasiliev,
  ``Higher Spin Algebras And Quantization On The Sphere And Hyperboloid,''
  Int.\ J.\ Mod.\ Phys.\ A {\bf 6}, 1115 (1991).



\bibitem{Vasiliev:1995dn}
M.~A.~Vasiliev,
``Higher Spin Gauge Theories in Four-Dimensions, Three-Dimensions, and Two-Dimensions,''
Int.\ J.\ Mod.\ Phys.\ D {\bf 5} (1996) 763
[hep-th/9611024].


\bibitem{Vasiliev:1999ba} 
  M.~A.~Vasiliev,
  ``Higher spin gauge theories: Star product and AdS space,''
  In *Shifman, M.A. (ed.): The many faces of the superworld* 533-610
  [hep-th/9910096].


  
\bibitem{Vasiliev:2000rn} 
  M.~A.~Vasiliev,
  ``Higher spin symmetries, star product and relativistic equations in AdS space,''
  hep-th/0002183.




\bibitem{Blencowe:1988gj} 
  M.~P.~Blencowe,
  ``A Consistent Interacting Massless Higher Spin Field Theory In D = (2+1),''
  Class.\ Quant.\ Grav.\  {\bf 6}, 443 (1989).

\bibitem{Bergshoeff:1989ns} 
  E.~Bergshoeff, M.~P.~Blencowe and K.~S.~Stelle,
  ``Area Preserving Diffeomorphisms And Higher Spin Algebra,''
  Commun.\ Math.\ Phys.\  {\bf 128}, 213 (1990).
  


\bibitem{Prokushkin:1998bq}
S.~F.~Prokushkin and M.~A.~Vasiliev,
``Higher Spin Gauge Interactions for Massive Matter Fields in 3-D AdS Space-Time,''
Nucl.\ Phys.\ B {\bf 545} (1999) 385
[hep-th/9806236].



\bibitem{Prokushkin:1998vn}
S.~Prokushkin and M.~A.~Vasiliev,
``3-D Higher Spin Gauge Theories with Matter,''
hep-th/9812242.




\bibitem{Fradkin:1990qk}
E.~S.~Fradkin and V.~Y.~.Linetsky,
``Supersymmetric Racah Basis, Family of Infinite Dimensional Superalgebras, SU(infinity+1|Infinity) and Related 2-D Models,''
Mod.\ Phys.\ Lett.\ A {\bf 6} (1991) 617.





\bibitem{Achucarro:1987vz}
A.~Achucarro and P.~K.~Townsend,
``A Chern-Simons Action for Three-Dimensional Anti-de~Sitter Supergravity Theories,''
Phys.\ Lett.\ B {\bf 180} (1986) 89.




\bibitem{Witten:1988hc}
E.~Witten,
``(2+1)-Dimensional Gravity as an Exactly Soluble System,''
Nucl.\ Phys.\ B {\bf 311} (1988) 46.



\bibitem{Castro:2011iw} 
  A.~Castro, R.~Gopakumar, M.~Gutperle and J.~Raeymaekers,
  ``Conical Defects in Higher Spin Theories,''
  JHEP {\bf 1202}, 096 (2012)
  [arXiv:1111.3381 [hep-th]].




\bibitem{Gutperle:2011kf}
M.~Gutperle and P.~Kraus,
``Higher Spin Black Holes,''
JHEP {\bf 1105} (2011) 022
[arXiv:1103.4304 [hep-th]].



\bibitem{Ammon:2011nk}
M.~Ammon, M.~Gutperle, P.~Kraus and E.~Perlmutter,
``Spacetime Geometry in Higher Spin Gravity,''
JHEP {\bf 1110} (2011) 053
[arXiv:1106.4788 [hep-th]].




\bibitem{Castro:2011fm}
A.~Castro, E.~Hijano, A.~Lepage-Jutier and A.~Maloney,
``Black Holes and Singularity Resolution in Higher Spin Gravity,''
JHEP {\bf 1201} (2012) 031
[arXiv:1110.4117 [hep-th]].



\bibitem{David:2012iu} 
  J.~R.~David, M.~Ferlaino and S.~P.~Kumar,
  ``Thermodynamics of higher spin black holes in 3D,''
  JHEP {\bf 1211}, 135 (2012)
  [arXiv:1210.0284 [hep-th]].

\bibitem{Chen:2012ba}
  B.~Chen, J.~Long and Y.~-N.~Wang,
  ``Phase Structure of Higher Spin Black Hole,''
  JHEP {\bf 1303} (2013) 017
  [arXiv:1212.6593 [hep-th]].

\bibitem{Ferlaino:2013vga} 
  M.~Ferlaino, T.~Hollowood and S.~P.~Kumar,
  ``Asymptotic symmetries and thermodynamics of higher spin black holes in AdS3,''
  arXiv:1305.2011 [hep-th].

   
\bibitem{deBoer:2013gz}
J.~de Boer and J.~I.~Jottar,
``Thermodynamics of Higher Spin Black Holes in AdS$_{3}$,''
arXiv:1302.0816 [hep-th].


\bibitem{Perez:2013xi} 
  A.~Perez, D.~Tempo and R.~Troncoso,
  ``Higher spin black hole entropy in three dimensions,''
  arXiv:1301.0847 [hep-th].

\bibitem{Perez:2012cf} 
  A.~Perez, D.~Tempo and R.~Troncoso,
  ``Higher spin gravity in 3D: black holes, global charges and thermodynamics,''
  arXiv:1207.2844 [hep-th].
 
 
\bibitem{Campoleoni:2010zq}
A.~Campoleoni, S.~Fredenhagen, S.~Pfenninger and S.~Theisen,
``Asymptotic Symmetries of Three-Dimensional Gravity Coupled to Higher-Spin Fields,''
JHEP {\bf 1011} (2010) 007
[arXiv:1008.4744 [hep-th]].



 
  
\bibitem{Gaberdiel:2012uj}
M.~R.~Gaberdiel and R.~Gopakumar,
``Minimal Model Holography,''
J.\ Phys.\ A {\bf 46} (2013) 214002
[arXiv:1207.6697 [hep-th]].

\bibitem{Ammon:2012wc}
M.~Ammon, M.~Gutperle, P.~Kraus and E.~Perlmutter,
``Black Holes in Three Dimensional Higher Spin Gravity: a Review,''
arXiv:1208.5182 [hep-th].


\bibitem{Banados:1998gg}
M.~Ba\~nados,
``Three-Dimensional Quantum Geometry and Black Holes,''
hep-th/9901148.


\bibitem{Henneaux:2010xg}
M.~Henneaux and S.~-J.~Rey,
``Nonlinear $W_{\infty}$ as Asymptotic Symmetry of Three-Dimensional Higher Spin Anti-de~Sitter Gravity,''
JHEP {\bf 1012} (2010) 007
[arXiv:1008.4579 [hep-th]].


\bibitem{Gaberdiel:2011wb}
M.~R.~Gaberdiel and T.~Hartman,
``Symmetries of Holographic Minimal Models,''
JHEP {\bf 1105} (2011) 031
[arXiv:1101.2910 [hep-th]].

\bibitem{Campoleoni:2011hg}
A.~Campoleoni, S.~Fredenhagen and S.~Pfenninger,
``Asymptotic W-Symmetries in Three-Dimensional Higher-Spin Gauge Theories,''
JHEP {\bf 1109} (2011) 113
[arXiv:1107.0290 [hep-th]].



\bibitem{Coussaert:1995zp}
O.~Coussaert, M.~Henneaux and P.~van Driel,
``The Asymptotic Dynamics of Three-Dimensional Einstein Gravity with a Negative Cosmological Constant,''
Class.\ Quant.\ Grav.\ {\bf 12} (1995) 2961
[gr-qc/9506019].





\bibitem{Drinfeld:1984qv}
V.~G.~Drinfeld and V.~V.~Sokolov,
``Lie Algebras and Equations of Korteweg-De Vries Type,''
J.\ Sov.\ Math.\ {\bf 30} (1984) 1975.


\bibitem{Pope:1989sr} 
  C.~N.~Pope, L.~J.~Romans and X.~Shen,
  ``W(infinity) And The Racah-wigner Algebra,''
  Nucl.\ Phys.\ B {\bf 339}, 191 (1990).




\bibitem{Tan:2012xi}
H.~S.~Tan,
``Exploring Three-Dimensional Higher-Spin Supergravity Based on $Sl(N|N-1)$ Chern-Simons Theories,''
JHEP {\bf 1211} (2012) 063
[arXiv:1208.2277 [hep-th]].

\bibitem{Datta:2012km}
S.~Datta and J.~R.~David,
``Supersymmetry of Classical Solutions in Chern-Simons Higher Spin Supergravity,''
JHEP {\bf 1301} (2013) 146
[arXiv:1208.3921 [hep-th]].




\bibitem{Hikida:2012eu}
Y.~Hikida,
``Conical Defects and ${\mathcal{N}}\!=2$ Higher Spin Holography,''
arXiv:1212.4124 [hep-th].




\bibitem{Chen:2013oxa}
B.~Chen, J.~Long and Y.~-N.~Wang,
``Conical Defects, Black Holes and Higher Spin (Super-)Symmetry,''
JHEP {\bf 1306} (2013) 025
[arXiv:1303.0109 [hep-th]].




\bibitem{Campoleoni:2013iha}
A.~Campoleoni and S.~Fredenhagen,
``On the Higher-Spin Charges of Conical Defects,''
arXiv:1307.3745 [hep-th].



\bibitem{Campoleoni:2013lma}
A.~Campoleoni, T.~Prochazka and J.~Raeymaekers,
``A Note on Conical Solutions in 3D Vasiliev Theory,''
JHEP {\bf 1305} (2013) 052
[arXiv:1303.0880 [hep-th]].




\bibitem{serre}
J.P.~Serre,
Chapter VII, ``A Course in Arithmetic," 
 Springer (1973). 




\bibitem{Kraus:2011ds}
P.~Kraus and E.~Perlmutter,
``Partition Functions of Higher Spin Black Holes and Their CFT Duals,''
JHEP {\bf 1111} (2011) 061
[arXiv:1108.2567 [hep-th]].


\bibitem{Gaberdiel:2012yb}
M.~R.~Gaberdiel, T.~Hartman and K.~Jin,
``Higher Spin Black Holes from CFT,''
JHEP {\bf 1204} (2012) 103
[arXiv:1203.0015 [hep-th]].



\bibitem{Gaberdiel:2013jca}
M.~R.~Gaberdiel, K.~Jin and E.~Perlmutter,
``Probing Higher Spin Black Holes from CFT,''
arXiv:1307.2221 [hep-th].


\bibitem{Regge:1974zd}
T.~Regge and C.~Teitelboim,
``Role of Surface Integrals in the Hamiltonian Formulation of General Relativity,''
Annals Phys.\ {\bf 88} (1974) 286.





\bibitem{Banados:2012ue} 
  M.~Banados, R.~Canto and S.~Theisen,
  ``The Action for higher spin black holes in three dimensions,''
  JHEP {\bf 1207}, 147 (2012)
  [arXiv:1204.5105 [hep-th]].



\bibitem{Campoleoni:2012hp}
A.~Campoleoni, S.~Fredenhagen, S.~Pfenninger and S.~Theisen,
``Towards Metric-Like Higher-Spin Gauge Theories in Three Dimensions,''
J.\ Phys.\ A {\bf 46} (2013) 214017
[arXiv:1208.1851 [hep-th]].


\bibitem{Compere:2013gja}
G.~Comp�re and W.~Song,
``W Symmetry and Integrability of Higher Spin Black Holes,''
arXiv:1306.0014 [hep-th].




\bibitem{Balog:1990mu}
J.~Balog, L.~Feher, L.~O'Raifeartaigh, P.~Forgacs and A.~Wipf,
``Toda Theory and W Algebra from a Gauged Wznw Point of View,''
Annals Phys.\ {\bf 203} (1990) 76.


\bibitem{Dijkgraaf:1996iy}
R.~Dijkgraaf,
``Chiral Deformations of Conformal Field Theories,''
Nucl.\ Phys.\ B {\bf 493} (1997) 588
[hep-th/9609022].





\bibitem{Kraus:2006wn} 
P.~Kraus,
``Lectures on black holes and the AdS$_3$/CFT$_2$ correspondence,''
Lect.\ Notes Phys.\  {\bf 755}, 193 (2008)
{\tt [arXiv:hep-th/0609074]}.

\bibitem{Witten:2007kt} 
  E.~Witten,
  ``Three-Dimensional Gravity Revisited,''
  arXiv:0706.3359 [hep-th].

 

\bibitem{Banerjee:2012aj}
S.~Banerjee, A.~Castro, S.~Hellerman, E.~Hijano, A.~Lepage-Jutier, A.~Maloney and S.~Shenker,
``Smoothed Transitions in Higher Spin AdS Gravity,''
Class.\ Quant.\ Grav.\ {\bf 30} (2013) 104001
[arXiv:1209.5396 [hep-th]].

\bibitem{quotient}
J.~Mei, R-x.~Miao, and E.D.~Skvotsov,
private communication.


\bibitem{Compere:2013nba}
G.~Comp\`ere, J.~I.~Jottar and W.~Song,
``Observables and Microscopic Entropy of Higher Spin Black Holes,''
arXiv:1308.2175 [hep-th].



\bibitem{Manschot:2007ha}
J.~Manschot and G.~W.~Moore,
``A Modern Farey Tail,''
Commun.\ Num.\ Theor.\ Phys.\ {\bf 4} (2010) 103
[arXiv:0712.0573 [hep-th]].








  


\end{thebibliography}
\end{document}